\documentclass[aps,prx,amsmath,amssymb,amsfonts,superscriptaddress,notitlepage,reprint,longbibliography,floatfix
]{revtex4-2}

\usepackage{graphicx}
\usepackage{dcolumn}
\usepackage{bm}
\usepackage{amssymb}
\usepackage{amsmath}
\usepackage[svgnames]{xcolor}
\usepackage[colorlinks=true,
            linkcolor=red,
            urlcolor=blue,
            citecolor=DarkGreen]{hyperref}
\usepackage[english]{babel}
\usepackage{braket}
\usepackage{microtype}
\usepackage{bbm}
\usepackage{tikz}
\usepackage{siunitx}
\usepackage{physics}
\usepackage{csquotes}
\usepackage{multirow}

\usepackage{float}
\makeatletter
\let\newfloat\newfloat@ltx
\makeatother
\usepackage{algorithm}
\usepackage{algpseudocode}

\usepackage[capitalise]{cleveref}

\begin{document}




\title{Deterministic generation of a 20-qubit two-dimensional photonic cluster state}

\author{James~O'Sullivan}
\thanks{These authors contributed equally}
\thanks{\\  james.osullivan@cea.fr (Current address: CEA Paris-Saclay, 91190 Gif-sur-Yvette, France),\\ kevin.reuer@phys.ethz.ch}

\author{Kevin~Reuer}
\thanks{These authors contributed equally}
\thanks{\\  james.osullivan@cea.fr (Current address: CEA Paris-Saclay, 91190 Gif-sur-Yvette, France),\\ kevin.reuer@phys.ethz.ch}

\author{Aleksandr~Grigorev}
\author{Xi~Dai}
\author{Alonso~Hern\'{a}ndez-Ant\'{o}n}

\affiliation{Department of Physics, ETH Zurich, CH-8093 Zurich, Switzerland}
\affiliation{Quantum Center, ETH Zurich, CH-8093 Zurich, Switzerland}

\author{Manuel~H.~Muñoz-Arias}
\affiliation{Institut Quantique and Département de Physique, Université de Sherbrooke, Sherbrooke, Quebec J1K 2R1, Canada}

\author{Christoph Hellings}
\author{Alexander Flasby}
\author{Dante Colao Zanuz}
\author{Jean-Claude~Besse}
\affiliation{Department of Physics, ETH Zurich, CH-8093 Zurich, Switzerland}
\affiliation{Quantum Center, ETH Zurich, CH-8093 Zurich, Switzerland}
\author{Alexandre~Blais}
\affiliation{Institut Quantique and Département de Physique, Université de Sherbrooke, Sherbrooke, Quebec J1K 2R1, Canada}
\author{Daniel~Malz}
\affiliation{Department of Mathematical Sciences, University of Copenhagen,
Universitetsparken 5, 2100 Copenhagen, Denmark}
\author{Christopher~Eichler}
\altaffiliation[Current address: ]{Physics Department, University of Erlangen-Nuremberg, Staudtstraße 5, 91058 Erlangen, Germany}

\author{Andreas~Wallraff}
\thanks{andreas.wallraff@phys.ethz.ch}
\affiliation{Department of Physics, ETH Zurich, CH-8093 Zurich, Switzerland}
\affiliation{Quantum Center, ETH Zurich, CH-8093 Zurich, Switzerland}

\date{\today}

\begin{abstract}
Multidimensional cluster states are a key resource for robust quantum communication~\cite{Duer2001, Briegel2001,Muralidharan2008}, measurement-based quantum computing~\cite{Raussendorf2001, Raussendorf2003} and quantum metrology~\cite{Friis2017,Shettell2020}.  Here, we present a device capable of emitting large-scale entangled microwave photonic states in a two dimensional ladder structure. The device consists of a pair of coupled superconducting transmon qubits which are each tuneably coupled to a common output waveguide. This architecture permits entanglement between each transmon and a deterministically emitted photonic qubit. By interleaving two-qubit gates with controlled photon emission, we generate $2 \times n$ grids of time- and frequency-multiplexed cluster states of itinerant microwave photons. We measure a signature of localizable entanglement across up to 20 photonic qubits. We expect the device architecture to be capable of generating a wide range of other tensor network states such as tree graph states, repeater states~\cite{Zhan2023} or the ground state of the toric code~\cite{Kitaev2003}, and to be readily scalable to generate larger and higher dimensional states.

\end{abstract}

\maketitle

\section{Introduction}

\begin{figure*}
    \centering
    \includegraphics[width=0.95\textwidth]{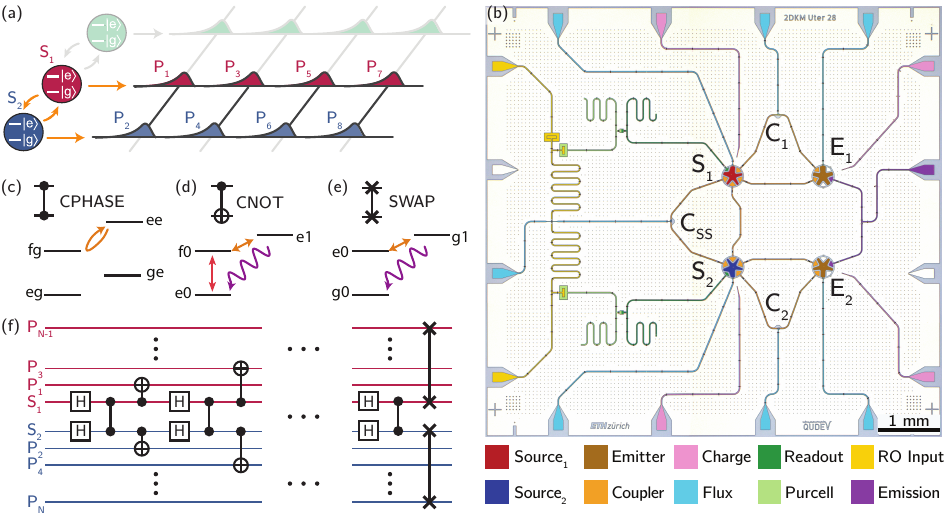}
    \caption{\textbf{A source of photonic tensor-network states} (a) Schematic of the entanglement and emission scheme. Couplings are depicted as orange arrows. (b) False-colour micrograph of the device, consisting of source qubits S$_1$ (red) and S$_2$ (blue), emitter qubits E$_1$ and E$_2$ (brown) and tunable couplers C$_1$, C$_2$ and C$_\mathrm{SS}$ (orange). Common readout (yellow) and emission (purple) lines permit frequency-multiplexed qubit readout and photon emission. (c) Illustration of the controlled-phase gate (CPHASE) between two source qubits by driving the $\ket{ee} \leftrightarrow \ket{fg}$ (orange) sideband transition. (d,e) Illustration of the physical implementation of (d) the controlled emission (CNOT) of a photon using a $\pi$ pulse on the $\ket{e}\leftrightarrow \ket{f}$ manifold (red), the $\ket{f0} \leftrightarrow \ket{e1}$ sideband transition (orange), and the  subsequent decay into the transmission line (purple), and (e) the emission of a photon via a SWAP operation using the $\ket{e0} \leftrightarrow \ket{g1}$ sideband transition (orange). (f) Quantum circuit used to generate a $2\times n$ cluster state. Source qubits are denoted as $\mathrm{S}_1$ (red), $\mathrm{S}_2$ (blue), emitted photonic qubits as $\mathrm{P}_i$. $\mathrm{H}$ denotes a Hadamard gate.}

    \label{fig:layout}
\end{figure*}


Multipartite entangled states, such as cluster states, are an essential resource for quantum computation and communication~\cite{Bennett2000, Horodecki2009}. Universal measurement-based quantum computing requires large-scale resource states with entanglement in at least two dimensions \cite{Raussendorf2001, Raussendorf2003}. Cluster states consisting of a few photonic qubits entangled in two dimensions have been generated using deterministic protocols and discrete variable encoding~\cite{Ferreira2024}, but for useful applications the scale of such states must be greatly increased. Generation of large 2D cluster states on-demand and with high fidelity remains an outstanding challenge.


State generation based on the heralding of successful detection events can achieve high fidelities. However, due to the probabilistic nature of heralding methods, the success rate decreases rapidly with increasing qubit number. Deterministic protocols are therefore strongly preferred for practical applications. Cluster states have been generated deterministically in matter-based~\cite{Cao2023, Lanyon2013,Friis2018,Wang2018,Wei2020} and photonic systems~\cite{Schwartz2016,Larsen2019,Asavanant2019,Istrati2020,Ferreira2024,Roh2023}. Tensor network states (the class of states to which cluster states belong to) in matter-based systems~\cite{Lanyon2013,Friis2018,Wang2018,Wei2020} are limited in size by the number of available stationary qubits. One way around this limitation is to reuse qubits, either via mid-circuit reset~\cite{Foss-Feig2022} or by repeatedly entangling and emitting itinerant photonic qubits~\cite{Schoen2005}. Emitting itinerant photonic qubits permits the generation of states consisting of far more photonic qubits than the number of available stationary qubits. Continuous-variable systems were used to generate large-scale photonic two-~\cite{Larsen2019,Asavanant2019} and three-dimensional cluster states~\cite{Roh2023}. However, states using discrete-variable encoding in both the microwave~\cite{Besse2020a,Ferreira2024} and the optical regime~\cite{Schwartz2016,Istrati2020} were limited so far to 10 photonic qubits for 1D entanglement~\cite{Besse2020a} and to 8 photonic qubits in 2D using post-selection~\cite{Thomas2024} (4 photonic qubits without post-selection~\cite{Ferreira2024}). Therefore, deterministic generation of significantly larger multidimensional tensor network states has yet to be demonstrated.

Superconducting circuits~\cite{Kjaergaard2020, Blais2021} are an excellent platform for realising devices capable of generating such multi-qubit cluster states~\cite{Besse2020a,Ferreira2024}. Superconducting qubits may be engineered to couple strongly to coplanar waveguide structures~\cite{Wallraff2004,Houck2007} and tunable couplers allow for rapidly switchable interactions between circuit elements with a high on/off ratio~\cite{Collodo2020}.

Here, we use a superconducting device to deterministically emit a two-dimensional cluster state consisting of itinerant microwave-frequency photonic qubits. By sequentially emitting time- and frequency multiplexed photonic qubits, we generate a $2\times n$ ladder-like cluster state, as illustrated in Fig.~\ref{fig:layout}(a). We then characterize the generated state using efficient quantum state tomography methods~\cite{Eichler2011, Baumgratz2013}. Our approach takes advantage of sequential emission of photonic qubits to reduce the required number of stationary qubits and potentially lossy elements such as memories~\cite{Naik2017, Matanin2023} or delay lines~\cite{Pichler2016,Ferreira2024}. This approach allows the system to emit states with a higher fidelity and consisting of significantly more qubits than the current state-of-the-art~\cite{Ferreira2024, Thomas2024}.

\section{Device architecture and cluster state generation protocol}
\label{sec:device_and_protocol}

To generate ladder-like cluster states, we use two transmon qubits whose lowest three levels we label $\ket{g}$, $\ket{e}$ and $\ket{f}$, tunably coupled to each other~\cite{Koch2007}. We refer to these as `source qubits', S$_1$ and S$_2$. In addition, each source qubit is tunably coupled to a waveguide, allowing for controlled emission of photons P$_i$, see Fig.~\ref{fig:layout}(a). To generate multipartite entangled states, $\mathrm{S}_1$ and $\mathrm{S}_2$ are entangled and subsequently emit itinerant photonic qubits P$_i$ (red and blue photon envelopes) sequentially. The procedure is repeated many times to generate $2 \times n$ ladder-like cluster states.
We implement this scheme with a device consisting of four superconducting transmon qubits as shown in Fig.\ref{fig:layout}(b). The source qubits  $\mathrm{S}_1$ (red) and $\mathrm{S}_2$ (blue) are transmon qubits with fixed $\ket{g}\leftrightarrow\ket{e}$ transition frequencies of 5.589~GHz (S$_1$, $T_1=\SI{27}{\micro\second}$, $T_2^*=\SI{22}{\micro\second}$) and 5.619~GHz (S$_2$, $T_1=\SI{22}{\micro\second}$, $T_2^*=\SI{23}{\micro\second}$). We realize the tunable coupling to a common waveguide for each source qubit via a frequency-tunable emitter qubit (brown, E$_1$ at 5.754~GHz and  E$_2$ at 5.354~GHz) and a tunable coupler (orange, $\mathrm{C}_1$ and $\mathrm{C}_2$).
The emitter qubits are strongly coupled to the common waveguide, such that any excitation of the emitter qubits rapidly decays into the waveguide as an itinerant photonic qubit with characteristic decay times of  $T_1 =2/\kappa= 54$~ns (E$_1$) and 42~ns (E$_2$).
The coupling between the source qubits is also realized by a tunable coupler, $\mathrm{C}_{\mathrm{SS}}$.  The tunable couplers consist of two parallel coplanar waveguides (CPWs) coupling two neighbouring qubits~\cite{Mundada2019,Collodo2020,Besse2020a}. One CPW has a superconducting quantum interference device (SQUID) placed in the middle of the centre conductor allowing flux-tuning of the CPW's inductance and thus its effective electrical length. Choosing a particular bias, we can utilize destructive interference between the fields propagating along each path to cancel the interaction between the two qubits~\cite{Mundada2019,Collodo2020,Besse2020a}. By modulating the flux threading the SQUID loop using a flux line, we parametrically drive the sideband transitions illustrated in Fig.~\ref{fig:layout}(c-e), enabling two-qubit gates~\cite{Reagor2018,Mundada2019,Besse2020a,Reuer2022}.


Figure~\ref{fig:layout}(f) shows the quantum circuit used to generate a $2\times n$ ladder-like cluster state. The protocol begins with a Hadamard gate on both $\mathrm{S}_1$ and $\mathrm{S}_2$ followed by a CPHASE gate between these two qubits. To implement the CPHASE gate, we parametrically drive a $2\pi$ rotation on the $\ket{ee} \leftrightarrow \ket{fg}$ transition~\cite{Reagor2018}, as illustrated in Fig.~\ref{fig:layout}(c). This imparts a conditional relative geometric phase between the two qubits~\cite{Strauch2003,Reagor2018}. The phase accumulated depends on the detuning of the $\ket{ee} \leftrightarrow \ket{fg}$ pulse~\cite{DiCarlo2010}---a resonant pulse results in a phase shift of $\pi$, thus enacting a CPHASE gate. The CPHASE gate is followed by controlled emission of a photonic qubit from $\mathrm{S}_1$ and $\mathrm{S}_2$ via CNOT gates. We perform two-qubit gates between a source qubit S$_1$ or S$_2$ and a photonic qubit P$_i$ by using the emitter qubits, whose lowest two levels we label $\ket{0}$ and $\ket{1}$. The CNOT gate, shown in Fig.~\ref{fig:layout}(d), is realized by first performing a $\pi_{ef}$ rotation on the source qubit and then parametrically driving the $\ket{f0} \leftrightarrow \ket{e1}$ transition~\cite{Besse2020a, Reuer2022}. If excited, the emitter subsequently decays into the waveguide, emitting an itinerant microwave photon~\cite{Houck2007,Eichler2011}. A photon is only emitted if the source qubit is initially in $\ket{e}$, realizing a CNOT gate between the source and the emitted photonic qubit. For the emission of the first $n-1$ pairs of photons a gate sequence consisting of Hadamard-CPHASE-CNOT is repeated $n-1$ times. In the last emission step,  emission of the photonic qubits is done via a SWAP gate from $\mathrm{S}_1$ to $\mathrm{E}_1$ and from $\mathrm{S}_2$ to $\mathrm{E}_2$ instead of a CNOT gate. We perform such a SWAP gate by transferring the excitation into the emitter qubit by parametrically driving the $\ket{e0} \leftrightarrow \ket{g1}$ transition~\cite{Besse2020a, Reuer2022}, as shown in Fig.~\ref{fig:layout}(e). These final SWAP gates disentangle S$_1$ and S$_2$ from the generated entangled photonic state.

To reduce cross-talk between the source qubits, which are detuned by less than 30 MHz (due to a frequency targeting error), we choose to perform single qubit gates on both S$_1$ and S$_2$ in 128~ns, slower than state-of-the-art~\cite{Werninghaus2021,Lazar2023}, thereby decreasing the spectral overlap between the drive pulses. For the two-qubit gates, we optimize gate times resulting in a 173~ns CPHASE gate. We perform the CNOT gates in 110~ns (S$_1$) and 106~ns (S$_2$), and the SWAP gates in 186~ns (S$_1$) and 240~ns  (S$_2$).
To ensure the emitter qubits fully decay to their ground state after a SWAP or CNOT gate, we choose a 650~ns delay between successive emission of photon pairs in our protocol. All relevant device parameters are summarized in App.~\ref{Appendix:experimental_setup}, Tab.~\ref{table:parameters}. Further details of the detection scheme and experimental setup are given in App.~\ref{Appendix:experimental_setup}.

\begin{figure*}
    \centering
    \includegraphics[width=0.99\textwidth]{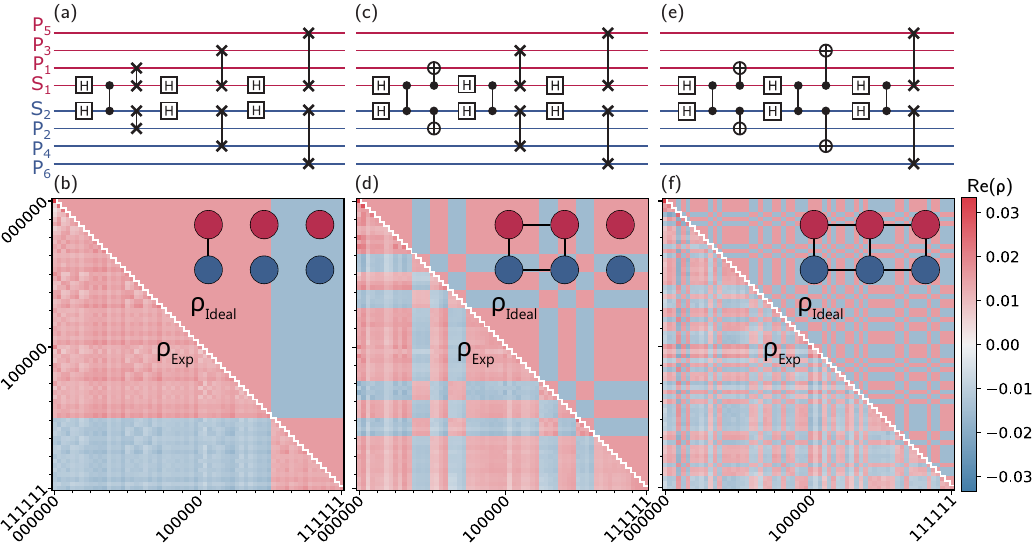}
    \caption{\textbf{Step by step build-up of a six-qubit cluster state.} We prepare six photonic qubits in three separate experiments, starting with a gate sequence containing only one entangling CPHASE gate (a),  four entangling gates (c),  and finally the full sequence to generate a six-qubit cluster state (e). The resulting reconstructed density matrices are shown in (b,d,f), below the corresponding gate sequences. For each density matrix an illustration of the entanglement structure is also depicted in the upper-right corner.}
    \label{fig:Cluster_appearance}
\end{figure*}

\begin{figure}
    \includegraphics[width=0.48\textwidth]{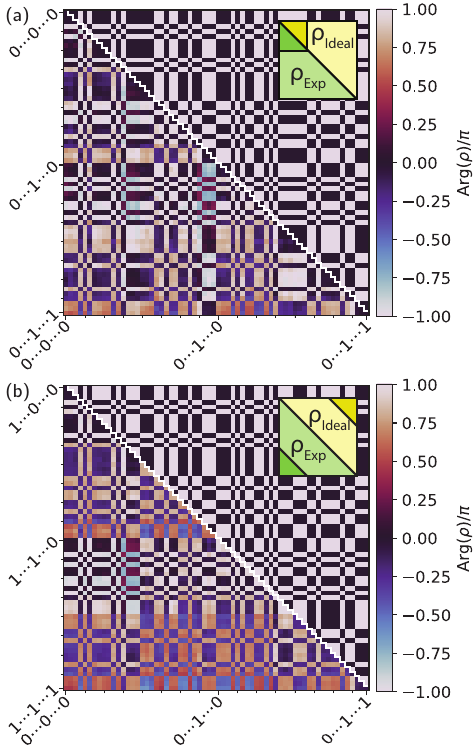}
    \caption{\textbf{Sections of the density matrix of a 20 photonic qubit state.} Phase of the experimental $\rho_{\text{Exp}}$ (below diagonal) and the ideal $\rho_{\text{Ideal}}$ (above diagonal) density matrix of the 20 qubit cluster state, showing only (a) the upper-left part and (b) far off-diagonal part of the full density matrix. Darker coloured triangles in the insets indicate the sections of the density matrix displayed.}
    \label{fig:rho_2d_states}
\end{figure}
\begin{figure}
    \centering
    \includegraphics[width=0.48\textwidth]{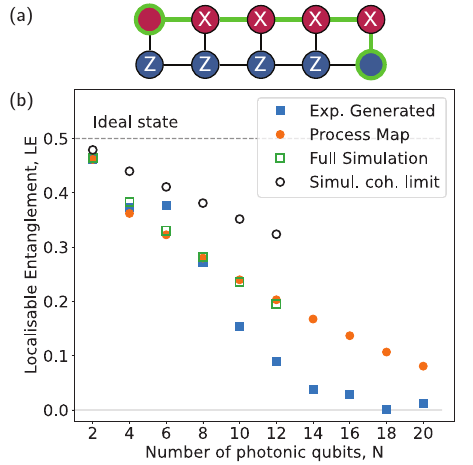}
    \caption{\textbf{Quantifying entanglement of the generated states.} (a) Schematic illustrating the chosen localizable entanglement metric for a 10 photonic qubit state. Qubits are projected into either the $X$ or the $Z$ bases, as indicated, then localizable entanglement is calculated for the two corner qubits (green edges). (b) Localizable entanglement as a function of number of generated photonic qubits $N= 2 \times n$, as measured by matrix product operator tomography of the experimentally generated states (exp. generated, blue, solid squares), as estimated from process maps (orange, solid circles), and as obtained by master equation simulations taking into account all errors (green, empty squares) and only decoherence (simul. coh. limit (simulated coherence limit), black, open circles), and the localizable entanglement of the ideal state (grey, dashed line). Shown is the mean over all possible paths of length $n$.}
    \label{fig:negativity_and_schematic}
\end{figure}

\section{State characterization}

When generating an entangled photonic state, the emitted photons P$_i$ propagate into the common waveguide and are routed off chip into a coaxial output line. We characterize the emitted photonic modes P$_i$ by recording their respective quadratures $I_i$ and $Q_i$~\cite{Eichler2012}. From $I_i$ and $Q_i$, we then calculate statistical moments $\langle \hat{a}^{\dagger s} \hat{a}^t \rangle$~\cite{Eichler2011} of the emitted photonic modes P$_i$. As we find that the second order correlation $\langle \hat{a}^{\dagger 2} \hat{a}^2 \rangle$ is close to zero (see App.~\ref{Appendix:Single_photon_characterisation} for details), we truncate the Hilbert space to $s,t \in \{0,1\}$, i.e.~the single photon Hilbert space. To then reconstruct the density matrix of up to 4-qubit states from the integrated quadratures, we calculate joint moments $\langle \hat{a}_1^{\dagger s} \hat{a}_1^{t} \hat{a}_2^{\dagger u} \hat{a}_2^{v} \hat{a}_3^{\dagger w} \hat{a}_3^{x} \hat{a}_4^{\dagger y} \hat{a}_4^{z} \rangle$, where $s,t... ,z \in \{0,1\}$, of all constituent photonic qubits P$_i$ and perform maximum likelihood estimation on the extracted moments to obtain the most likely physical density matrix~\cite{Eichler2012,Besse2020a}, see App.~\ref{Appendix:MPS_tomography} for details. 

As the number of required measurement runs to achieve a given signal-to-noise ratio $k \propto 1/\eta^O$, where $\eta$ is the quantum efficiency of the detection chain, scales exponentially with the order $O$ of the moment~\cite{daSilva2010}, measuring joint moments of more than four photonic qubits becomes challenging.
Thus, for states consisting of more than four photonic qubits, we use a two-step reconstruction process. Qubits in the generated states have at most 3 nearest neighbours. We reconstruct the density matrices of all four-qubit subsets of the generated state consisting of each qubit and its nearest neighbours. From these we efficiently reconstruct a matrix product operator representing the full density matrix, as described in Ref.~\cite{Baumgratz2013}, making use of the fact that the cluster state is the unique ground state of a gapped nearest-neighbour Hamiltonian~\cite{Briegel2001}. The algorithm iteratively finds the most likely state consistent with the reduced density matrices, see App.~\ref{Appendix:MPS_tomography} for details.


To illustrate the generation of the cluster states step by step, and to demonstrate the flexibility of the device in generating different states, we generate three states each consisting of six photons, each using a different number of entangling gates. The quantum circuits used to generate these three states are shown in Figs.~\ref{fig:Cluster_appearance}(a,c,e). In Figs.~\ref{fig:Cluster_appearance}(b,d,f), we show the reconstructed density matrices of these states, whose entanglement structure is illustrated in the insets. In (a,b), we perform only a single CPHASE entangling gate in the first emission step, followed by SWAP gates. In the other emission steps, we only apply Hadamard and SWAP gates, thus not creating entanglement. In (c,d), we add additional CHPASE and CNOT gates to create a four-qubit cluster state in a six-qubit subspace. Finally, in (e,f), we perform the full sequence to generate a six-qubit cluster state. In each step we see that the addition of more entangling gates introduces additional structure to the density matrix, resulting in the intricate pattern of stripes in Fig.~\ref{fig:Cluster_appearance}(f) that corresponds to a six-qubit cluster state.

We performed the $2 \times n$-qubit cluster state generation protocol outlined above for $n \in \{2, 3, 4, \ldots, 10\}$, resulting in cluster states of up to 20 photonic qubits. We find that the reconstructed $N$-photon cluster state density matrices $\rho_\text{Exp}$ have fidelities $F=\Tr \left( \sqrt{\rho_\text{Exp} \rho_\text{Ideal}} \right)^2$ of 0.84 (four qubit), 0.77 (six qubit), 0.59 (eight qubit) and 0.12 (20 qubit) to the corresponding ideal cluster state $\rho_\text{Ideal}$ (see App.~\ref{Appendix:MPS_tomography} for more details). For the four, six and eight qubit states, the fidelity therefore exceeds the threshold for genuine multipartite entanglement of 50\%, i.e. the generated state cannot be written as a convex sum of bi-separable states~\cite{Guhne2009}.


The density matrix of the 20 qubit cluster state, containing a total of $2^{40} \approx 1 \times 10^{12}$ entries, is too large to display and impractical to construct as a density matrix in full, so we instead reconstruct the state entirely in the matrix product operator representation and only show the upper-left part and far off-diagonal part of the density matrix, see Fig.~\ref{fig:rho_2d_states}(a,b) for the phase and App.~\ref{Appendix:MPS_tomography} for the magnitude. We observe that for all cluster state density matrices for $N=2,4,...,20$, including those not shown, elements closer to the $|00...\rangle$ state are larger in magnitude compared to the ideal state while far off-diagonal elements are smaller in magnitude than the ideal state. This is due to source qubit decay and dephasing occurring during the protocol (see App.~\ref{Appendix:MPS_tomography}). As this dephasing causes the diagonal elements of $\rm{Re}(\rho)$ to be much larger than the off-diagonal elements, we instead plot the phase in Fig.~\ref{fig:rho_2d_states}(a,b), which allows us to see the pattern due to entanglement more clearly. 

To further characterize the generated multi-qubit states, we estimate up to which number of qubits measurable entanglement persists across the multi-qubit state. For states whose fidelity is less than 50\% we must use an another metric than fidelity to determine if entanglement persists. We therefore calculate the localizable entanglement between the first P$_1$ and last P$_N$ photonic qubit~\cite{Hein2006}. To do so, we project all other qubits either in the $X$ or the $Z$ basis and then evaluate the negativity between P$_1$ and P$_N$. For a cluster state a $Z$ projection removes the node and its bonds, while an $X$ projection preserves the entanglement bonds~\cite{Hein2006}. We apply these projections to the reconstructed state in post-processing. Using $X$ projections, we construct a path of length $n=N/2$, the shortest possible distance, between P$_1$ and P$_N$, while applying $Z$ projections on all nodes outside the path; see App.~\ref{appendix:LE} for details. Figure~\ref{fig:negativity_and_schematic}(a) illustrates an example path for $N=10$. As each projection has a probabilistic outcome, for an $N$-qubit state, there are $2^{N-2}$ possible outcomes for each path. For states larger than $N=12$ it becomes impractical to compute all such outcomes. We therefore average the obtained negativity over 1024 randomly sampled projection outcomes and obtain the localizable entanglement for each path, then average over all paths with length $n=N/2$. A nonzero localizable entanglement value indicates that there is measurable entanglement between P$_1$ and P$_N$. 

We calculate the localizable entanglement for experimentally generated and  reconstructed states. In addition, we measure process maps of the two repeated processes (H + CPHASE + CNOT and H + CPHASE + SWAP) using process tomography. We then repeatedly apply the extracted process maps to the initial ground state density matrix $\ket{gg}\bra{gg}$ of the source qubits to calculate the expected density matrix of the generated cluster states~\cite{Schwartz2016,Besse2020a}, see App.~\ref{appendix:process_maps} for further details. From these density matrices we can also extract the localizable entanglement for comparison. Finally, we perform two types of master equation simulation of the state generation sequence, one taking into account only decay and dephasing errors, estimated from measured source qubit $T_1, T_2^*$ values, and one taking into account all errors (i.e. including coherent gate errors), see App.~\ref{appendix:simulation_details} for further details. From these simulations, we extract the localizable entanglement from the resulting simulated density matrices. The localizable entanglement values calculated from all above methods for states up to 20 photons in size are shown in Fig.~\ref{fig:negativity_and_schematic}. We see that the localizable entanglement decays approximately exponentially with state size in all curves. This is expected as larger system sizes involve longer gate sequences and therefore more decoherence and gate errors. The simulations including only decay and dephasing have larger localizable entanglement than those taking into account all errors, indicating that coherent gate errors make a significant contribution to the overall error in state generation. The localizable entanglement extracted from process maps agrees well with the simulation including all errors, while the experimentally generated and reconstructed density matrices for $N>8$ yield localizable entanglement values slightly below that of the process maps and simulations. This is most likely due to errors introduced by the reconstruction method. To verify this, we apply the reconstruction method to density matrices obtained from master equation simulations and observe a reduction in the extracted localizable entanglement and fidelity of the reconstructed density matrices compared to the simulated density matrices, which do not require  reconstruction, see App.~\ref{Appendix:MPS_tomography} for details. The reconstruction method thus, at least for the simulated states, results in a conservative estimate of the localizable entanglement. Using this conservative estimate, we see that the localizable entanglement is nonzero in the reconstructed 20 qubit state and is predicted by the recorded process maps to persist in even larger states.

From the analysis of process maps and master equation simulations, we conclude that the ultimate size of the states we can generate is limited both by the fidelity of our single and two-qubit gates and from decoherence during the pulse sequence. Qubit drive-line crosstalk is likely primarily responsible for the single qubit gate error and crosstalk cancellation should alleviate this. Further optimization of the coupler design and qubit frequencies and fine-tuning of the gate parameters may be feasible to increase two-qubit gate (in particular, the CPHASE gate) fidelities, which we, using simulations, estimate to currently be around 97~\%, towards the state-of-the-art performance for similar tunable coupler designs~\cite{Marxer2023}. Extending qubit coherence times from \SI{22}{\micro\second} (S$_1$) and \SI{23}{\micro\second} (S$_2$) towards the state of the art, which is more than an order of magnitude longer~\cite{Wang2022}, will reduce the decoherence-related error that we observe in the largest states generated. This will permit the utilization of significantly longer gate sequences to generate larger 2D entangled states.

\section{Conclusion}
In conclusion, we demonstrate deterministic generation of cluster states consisting of up to 20 photons entangled in two dimensions with nonzero localizable entanglement persisting across 20 constituent photons. We extract a fidelity of 84\%, 77\% and 59\% for four-, six- and eight-photon 2D entangled states, respectively. We employ efficient tomography techniques to reconstruct the density matrix of states whose size precludes the use of direct
tomography. These techniques are not specific to superconducting circuits and could be applied in the context of any physical platform. We perform process tomography on the constituent parts of the experimental gate sequence and master equation simulations to extract the expected quality of generated states. 

With the previously outlined improvements to our device, it should be feasible to significantly extend the duration of our protocol to generate larger cluster states. In addition, our architecture is scalable by fabricating additional stationary qubits. Two-qubit gates could be alternated such that all such gates necessary for an $m\times n$ sized 2D-entangled cluster state could be performed in no more than two sequential steps for any number $m$ of source qubits, making the scaling of gate sequence time in this protocol extremely attractive. 

The device architecture is in principle capable of creating states such as repeater states or tree graph states, which are relevant for a variety of quantum communication protocols~\cite{Borregaard2020, Bell2023}. Addition of either a quantum memory or greater connectivity between qubits could allow generation of higher-dimensional states. The prospect to generate a wide variety of entangled photonic states at a significantly larger scale opens up exciting possibilities for using such states for measurement-based quantum computing, metrology or quantum communication protocols in the context of the fast-growing field of waveguide QED.

\section*{Data availability}
The data supporting the findings of this article is available upon request.

\section*{Competing interests}
The authors declare no competing interests.

\section*{Acknowledgements}
We are grateful for the contributions of Alexander Ferk to the experimental setup and Graham J. Norris for photomask fabrication. This work was supported by the Swiss National Science Foundation (SNSF) through the project ``Quantum Photonics with Microwaves in Superconducting Circuits'' (Grant No. 200021\_184686), by the European Research Council (ERC) through the project ``Superconducting Quantum Network'' (SuperQuNet) and by ETH Zurich. Support is also acknowledged from the U.S. Department of Energy, Office of Science, National Quantum Information Science Research Centers, Quantum Systems Accelerator. Additional support is acknowledged from  the Ministère de l’Économie et de l’Innovation du Québec, NSERC, and the Canada First Research Excellence Fund. D.M. acknowledges financial support from the Novo Nordisk Foundation under grant numbers NF22OC0071934 and NNF20OC0059939.

\section*{Author Contributions}
J.O. and K.R. designed the device. A.F., D.C.Z. and J.-C.B. fabricated the device. J.O. and K.R. prepared the experimental setup. J.O., K.R., A.G., X.D. and A.H. characterized and calibrated the device and the experimental setup. J.O, K.R., A.G., X.D. and A.H. carried out the experiments and analyzed the data, with support from D.M. and C.H.. M.M. performed the master equation simulations. J.O. and K.R. wrote the manuscript with input from all co-authors. J.-C.B., A.B., C.E. and A.W. supervised the work.

\begin{appendix}

\section{Experimental setup and device calibration}
\label{Appendix:experimental_setup}

\begin{figure*}
    \centering
    \includegraphics[width=0.95\textwidth]{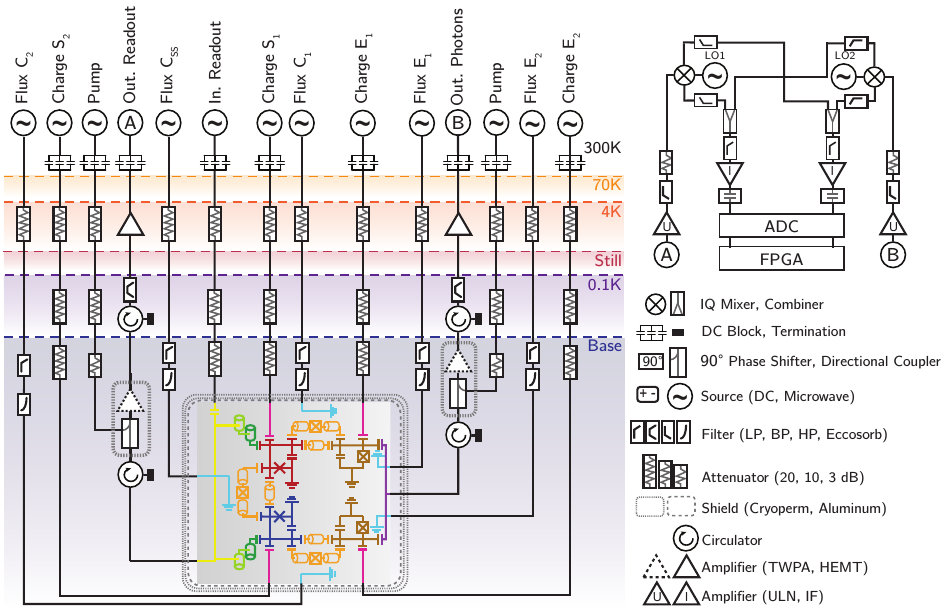}
    \caption{\textbf{Setup.} Schematic of the cryogenic and room temperature wiring.}
    \label{fig:appendix_setup}
\end{figure*}

The device shown in Fig.~\ref{fig:layout}(a) is mounted at the base plate of a dilution refrigerator at approximately $30~$mK. The sample is placed inside a superconducting aluminium shield which, in turn, is placed within two Cryophy magnetic shields, see sketch in Fig.~\ref{fig:appendix_setup}. The outermost shield forms a light-tight seal with a copper lid. Microwave drive lines are attenuated by 20 dB each on the 4~K, 100~mK, and base plate~\cite{Krinner2019}. The four qubits each have a microwave drive line, the readout resonators are driven via a single common microwave drive line. The output lines for readout and photon emission each are shared and are each amplified by a travelling wave parametric amplifier (TWPA) followed by a high electron mobility transistor (HEMT) amplifier at 4~K and further amplification at room temperature. The output signals are down-converted to an intermediate frequency with an $IQ$ mixer, and subsequently each quadrature ($I$ and $Q$) is digitized separately. As we use a digitizer with only two input channels, we combine the $I$ components from both output lines using high- and low-pass filters and route them into one input channel; we do the same for the $Q$ component.
To bias the tunable coupler and emitter qubits and to activate two-qubit gates, we use a total of 5 flux lines. A full diagram of the wiring of the sample is shown in Fig.~\ref{fig:appendix_setup}.

\begin{figure}
    \centering
    \includegraphics[width=0.90\columnwidth]{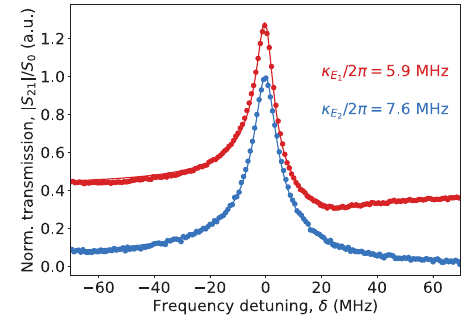}
    \caption{\textbf{Emitter qubit spectroscopy.} Absolute value of transmission coefficient $|S_{21}|/S_0$ vs detuning $\delta$ from the emitter qubit, normalized to the transmission coefficient $S_0$ at $\delta=0$ when driving the respective emitter qubit through its own charge line. Lines are fits to the data, from which we extract the indicated linewidths. The transmission coefficient for $\mathrm{E}_1$ is offset by 0.3 for better visibility.}
    \label{fig:appendix_Spec}
\end{figure}

We measure the basic device parameters, see Table.~\ref{table:parameters}, using single- and two-tone spectroscopy, as well as Rabi, Ramsey and coherence measurements, as presented in Ref.~\cite{Besse2020a}. To characterize the frequency and linewidth of the emitter qubits, we measure the elastic scattering of a weak input tone on the emitter qubit into the photon output line. We then fit the Fano resonance \cite{Fano1961} function

\begin{align}
|S_{21}| &=\sqrt{\left|\frac{S_0}{1+q^2} \frac{(\epsilon+q)^2} {\epsilon^2+1}\right|} \text{ with} \\
\epsilon &= \frac{4\pi\delta}{\kappa}
\end{align}

to the data, where $\delta$ is the detuning from the emitter qubit, $\kappa$ is the linewidth of the emitter, $q$ is a phenomenological shape parameter and $S_0$ is the transmission at $\delta=0$, see Fig.~\ref{fig:appendix_Spec}. The Fano resonance likely originates from crosstalk of the emitter qubit's charge line into the waveguide.

\begin{table}[!b]
\centering
\begin{ruledtabular}
 \begin{tabular}{l l r r}
 & & S$_1$/E$_1$ & S$_2$/E$_2$ \\ 
 \hline
 Source & $g$-$e$ frequency, $\omega_{ge}/2\pi$ [GHz] & 5.589 & 5.619 \\
 & $e$-$f$ frequency, $\omega_{ef}/2\pi$ [GHz] & 5.413 &  5.438 \\
 & anharmonicity, $\alpha/2\pi$ [MHz] & 176 & 181 \\
 & lifetime of $\ket{e}$, $T_1^{(e)}$ [\SI{}{\micro\second}]& 27 & 22 \\
 & lifetime of $\ket{f}$, $T_1^{(f)}$ [\SI{}{\micro\second}]& 16 & 4 \\
 & dephasing time of $\ket{e}$, $T_2^{\star(ge)}$ [\SI{}{\micro\second}] & 22 & 23 \\
 & dephasing time of $\ket{f}$, $T_2^{\star(ef)}$ [\SI{}{\micro\second}] & 12 & 6 \\
 Emitter & $0$-$1$ frequency, $\omega_{01}/2\pi$ [GHz] & 5.754 & 5.354 \\
 & decay rate, $\kappa/2\pi$ [MHz] & 5.9 & 7.6 \\
 Coupler & $\ket{e0} \leftrightarrow \ket{g1}$ coupling, $J/2\pi$ [MHz] & 2.5  & 2.5 \\
 & $\ket{f0} \leftrightarrow \ket{e1}$ coupling, $J/2\pi$ [MHz] & 3.4  & 3.6 \\
 S$_1$S$_2$ coupl. & $\ket{fg} \leftrightarrow \ket{ee}$ coupling, $J/2\pi$ [MHz] & \multicolumn{2}{c}{2.9} \\
 \end{tabular}
\end{ruledtabular}
 \caption{\label{table:parameters}Measured device parameters.}
\end{table}

\begin{figure}
    \centering
    \includegraphics[width=0.90\columnwidth]{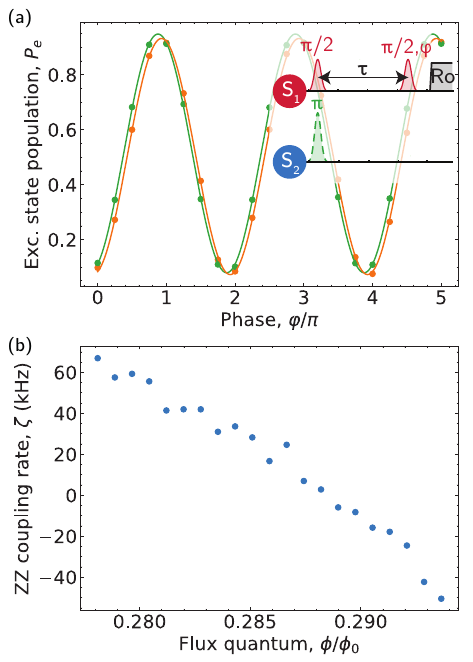}
    \caption{\textbf{Static ZZ cancellation.}(a) Excited state population $P_e$ of $\mathrm{S}_1$ after a Ramsey-like experiment (see inset) vs phase of the second $\pi/2$ pulse for $\mathrm{S}_2$ in the ground (orange) or excited state (green). Fits to the data are shown as solid lines. (b) Measured ZZ coupling rate $\zeta$ vs flux quanta $\Phi/\Phi_0$ for the tunable coupler $C_{SS}$ around the turn off point.}
    \label{fig:staticZZ}
\end{figure}

To measure the $ZZ$ interaction rate $\zeta$ between the source qubits $\mathrm{S}_1$ and $\mathrm{S}_2$, we perform a Ramsey measurement on $\mathrm{S}_1$ while $\mathrm{S}_2$ is prepared either in the ground or the excited state. For the measurement on $\mathrm{S}_1$, we apply a $\pi/2$ pulse, wait for a fixed time $\tau$, and apply a second $\pi/2$ pulse with varying phase $\varphi$. The measured excited state population then follows a cosine curve with a certain offset phase $\phi_0$, see Fig.~\ref{fig:staticZZ}(a). This offset phase $\phi_0$ depends on the frequency of $\mathrm{S}_1$. If $\mathrm{S}_2$ is in the excited state, the frequency of $\mathrm{S}_1$ will be shifted by $\zeta$ due to the $ZZ$ interaction, therefore the offset phase will change. The $ZZ$ interaction rate $\zeta$ is then obtained from the difference between the phase offset $\phi_{0,g}$ when $\mathrm{S}_2$ is in the ground and $\phi_{0,e}$ when is $\mathrm{S}_2$ is in the excited state: $\zeta= (\phi_{0,g}-\phi_{0,e})/\tau$. The $ZZ$ interaction rate depends on the flux bias on the tunable coupler $\mathrm{C}_\text{SS}$ between the source qubits. We measure the $ZZ$ interaction rate at different bias points and choose a bias point where $\zeta<10$~kHz, see Fig.~\ref{fig:staticZZ}(b).

\begin{figure*}
    \centering
    \includegraphics[width=0.90\textwidth]{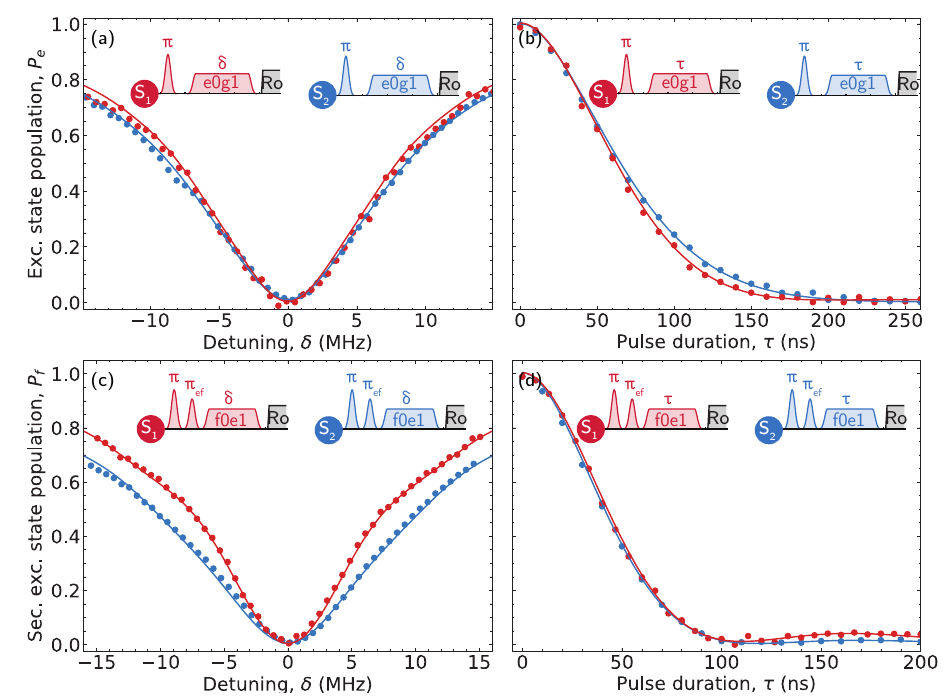}
    \caption{\textbf{Two-qubit gate calibrations}. (a,b) Measured excited state population $P_e$ of $\mathrm{S}_1$ (red) and $\mathrm{S}_2$ (blue) after preparing the respective source qubit in the excited state and applying a modulated flux pulse of duration $\tau$ and detuning $\delta$ $\ket{e0} \leftrightarrow \ket{g1}$ transition  to the respective tunable coupler (see insets) vs (a) detuning $\delta$ for a fixed duration $\tau=210$~ns (S$_1$)/$200$~ns (S$_2$) and (b) duration $\tau$ for a fixed detuning $\delta=0.5$~MHz (S$_1$)/$\delta=0.3$~MHz (S$_2$). (c,d) Measured second excited state population $P_f$ of $\mathrm{S}_1$ (red) and $\mathrm{S}_2$ (blue) after preparing the respective source qubit in the second excited state $\ket{f}$ and applying a modulated flux pulse of duration $\tau$ and detuning $\delta$ from the $\ket{f0} \leftrightarrow \ket{e1}$ transition to the respective tunable coupler (see insets) vs (c) detuning $\delta$ for a fixed duration $\tau=113$~ns (S$_1$), $159$~ns (S$_2$) and (d) duration $\tau$ for a fixed detuning $\delta=0.7$~MHz (S$_1$), $\delta=0.2$~MHz (S$_2$). Solid lines in (a-d) are fits to the data.}
    \label{fig:Chevrons}
\end{figure*}

To calibrate the SWAP gate between source and emitter qubits, we excite the source qubit into the first excited state, apply a flux pulse with a varying modulation frequency $\omega$ around the expected $\ket{e0}\leftrightarrow \ket{g1}$ sideband frequency and varying duration $\tau$, and read out the state of the source qubit afterwards. A $\pi$ pulse on the $\ket{e0}\leftrightarrow \ket{g1}$ transition will swap the excitation from the source to the emitter qubit. Thus, we can find the optimal modulation frequency $\omega$ and duration $\tau$ by minimizing the excited state population of the source qubit, see Fig~\ref{fig:Chevrons} (a) and (b). Similarly, we can find the parameters of a $\pi$ pulse on the $\ket{f0}\leftrightarrow \ket{e1}$ transition for the CNOT gate: We prepare the source qubit in the second excited state $\ket{f}$, apply a flux pulse of varying modulation frequency $\omega$ (around the expected $\ket{f0}\leftrightarrow \ket{e1}$ sideband frequency) and duration $\tau$, and read out the source qubit. For the optimal $\omega$ and $\tau$, the second excited state population is minimal, see Fig~\ref{fig:Chevrons} (c) and (d). The amplitude of the flux pulse is for both sideband transition optimized heuristically: Increasing the flux pulse amplitude leads to a shorter gate duration, but beyond a certain amplitude, gate fidelity drops off rapidly, possibly due to nonlinearities in the coupler caused by the large drive amplitude. Hence, we choose the maximum amplitude for which we observe no drop off in gate fidelity.

\begin{figure}
    \centering
    \includegraphics[width=0.90\columnwidth]{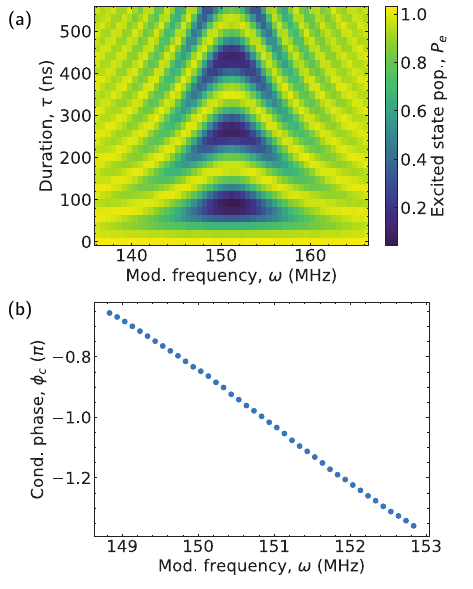}
    \caption{\textbf{CPHASE calibration} (a) Excited state population $P_e$ of $S_2$ after preparing $\mathrm{S}_1$ and $\mathrm{S}_2$ in the state $\ket{ee}$ and applying a modulated flux pulse of duration $\tau$ and modulation frequency $\omega$. (b) Extracted conditional phase $\phi_c$ vs modulation frequency $\omega$ of the flux pulse for a fixed duration $\tau=100$~ns.}
    \label{fig:CPHASE}
\end{figure}

To calibrate the CPHASE gate, we first calibrate the duration $\tau$ and modulation frequency $\omega$ of a $2 \pi$ rotation in the $\ket{fg}\leftrightarrow \ket{ee}$ sideband rotation. For this purpose, we prepare the source qubits in the state $\ket{ee}$ and then apply a flux pulse with varying $\tau$ and $\omega$ with a subsequent readout of $S_2$, similar to the SWAP and CNOT gate calibrations. We then optimize these parameters for optimal return into the $\ket{ee}$ state, i.e. maximizing the excited state population for the measured source qubit, see Fig.~\ref{fig:CPHASE}(a). We calibrate the conditional phase of the CPHASE gate by performing the measurement of the $ZZ$ interaction rate $\zeta$ described above, but apply an $\ket{ee}\leftrightarrow \ket{fg}$ pulse for the calibrated duration $\tau$ and modulation frequency $\omega$ in-between the two $\pi/2$ pulses. For our gate sequences, we aim for a conditional phase $\phi_c=\pi$, where the conditional phase is defined as $\phi_c=\phi_{0,g}-\phi_{0,e}$. The conditional phase $\phi_c$ depends on the modulation frequency $\omega$, so we adjust $\omega$ such that $\phi_c=\pi$, see Fig.~\ref{fig:CPHASE}(b). The optimal duration $\tau$ to return into the $\ket{ee}$ also depends on $\omega$. We therefore repeat the two measurements, until we converge at a $\tau$ and $\omega$ where both the conditional phase $\phi_c=\pi$ and the population of $\ket{ee}$ is maximized.

\section{Frequency multiplexing and mode matched filtering}
\label{Appendix:Photon_transients}

\begin{figure}
    \centering
    \includegraphics[width=0.90\columnwidth]{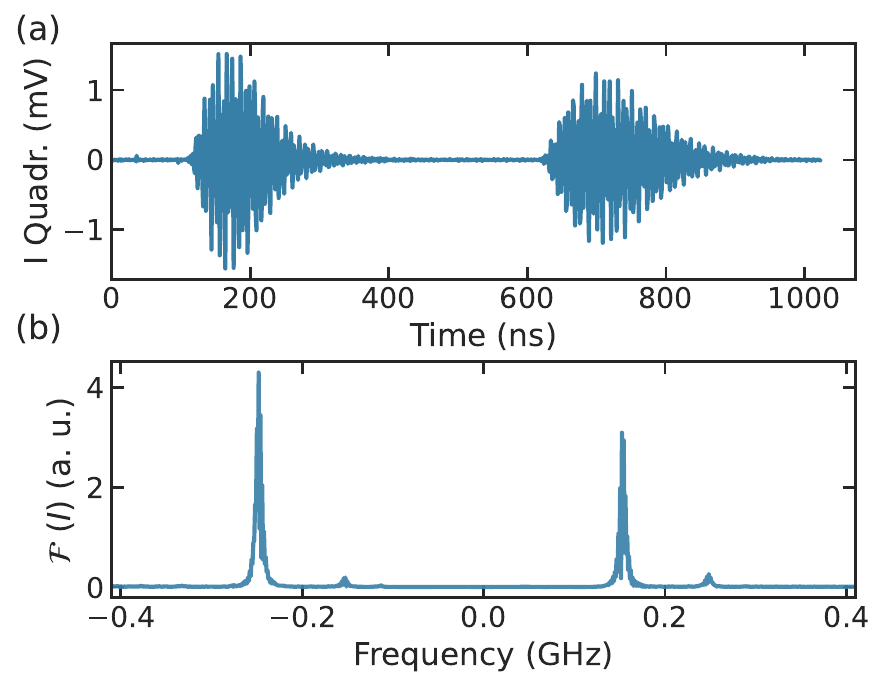}
    \caption{\textbf{Time and frequency multiplexing of photonic qubits.} (a) Time transient I quadrature signal of four emitted photons, multiplexed in time and frequency, after frequency down-conversion to an intermediate frequency, which is visible as a sinusoidal modulation of the signal. Pairs of photons are emitted simultaneously, hence only two envelopes are visible in time. (b) Fourier transform $\mathcal{F}(I)$ of the same quadrature signal, showing the presence of two distinct frequency peaks due to frequency multiplexing, corresponding to the frequencies of $\mathrm{E}_1$ and $\mathrm{E}_2$ after downconversion to an intermediate frequency.}
    \label{fig:Photon_transients}
\end{figure}

To test our emission protocol, we generate a superposition state of four photonic qubits arranged in a $2 \times 2$ grid. Specifically, we emit a photon in the state $\frac{1}{\sqrt{2}}\left(\ket{0}+\ket{1}\right)$ on both emitters via the $\ket{f0} \leftrightarrow \ket{e1}$ transition, followed by a second emission of the same state from each emitter via the $\ket{e0} \leftrightarrow \ket{g1}$ transition.
We then measure the amplified output field $a_{\text{out}}(t)$ from the emission line via heterodyne detection using the amplification chain and frequency downconversion outlined in Sec.~\ref{Appendix:experimental_setup}, see Fig.~\ref{fig:Photon_transients} for the measured $I$ quadrature.

We observe two frequency-modulated transient peaks, defining two-time bins. A Fourier-transform of each time bin reveals two transient peaks in frequency, defining two frequency bins. Note that the signal has been downconverted to match the range of our digitzer (-500~MHz to +500~MHz).


We use the transients measured from $\frac{1}{\sqrt{2}}\left(\ket{0}+\ket{1}\right)$ photonic states to define the mode of the photon. Therefore, by using the transient as an integration filter, we can maximize the detection efficiency~\cite{Eichler2012,Besse2020a}. Using such a time-dependent integration filter, we can easily distinguish between photonic qubits in different time bins. These integration filters also down-convert the signal from the intermediate frequency and filter out all other frequencies due to the sinusoidal modulation in $I$ and $Q$ of the photon transients which are used as the filter functions for each photonic qubit. Hence, the integration also distinguishes between photonic qubits of different frequencies, provided the photon transients have no overlap in frequency.

\section{Characterizing single photon emission}
\label{Appendix:Single_photon_characterisation}

By measuring the complex integrated time transient $\hat{S}=\hat{a}+\hat{h}^{\dagger}$, we directly probe the Husimi-Q function $Q_{\hat{S}(\alpha)}$ of the output signal. This consists of the mode $\hat{a}$ of the photonic qubit we wish to detect, convolved with Gaussian noise $\hat{h}^{\dagger}$ originating from vacuum fluctuations, attenuation in the output line and noise added by the amplification chain~\cite{Eichler2012}. The moments of the signal $\hat{S}$ are defined as:

\begin{align}
\langle (\hat{S}^{\dagger})^{t} \hat{S}^{s} \rangle &= \int_\alpha \alpha^{*,s} \alpha^t Q_{\hat{S}(\alpha)}
\end{align}

If we send no photonic qubit, we measure the Husimi-Q funtion of the noise mode $\hat{h}^{\dagger}$ of the system. We can then also extract the moments of the noise mode:
\begin{align}
\langle \hat{h}^{s} (\hat{h}^{\dagger})^{t}  \rangle &= \int_\alpha \alpha^{*,s} \alpha^t Q_{\hat{h^{\dagger}}(\alpha)}
\end{align}

To obtain the moments $\langle (a^{\dagger})^{i} a^{j} \rangle$ of the photonic field, we use that $\hat{S}=\hat{a}+\hat{h}^{\dagger}$, and therefore

\begin{equation}
\left\langle (\hat{S}^\dagger)^{t}\hat{S}^{s}\right\rangle _{p_{a}} = \sum _{s,t=0}^{t,s}\binom{s}{j}\binom{t}{i}\langle (a^{\dagger})^{i} a^{j} \rangle\langle h^{t-i} (h^{\dagger})^{s-j} \rangle
\end{equation}

\begin{figure*}
    \centering
    \includegraphics[width=0.90\textwidth]{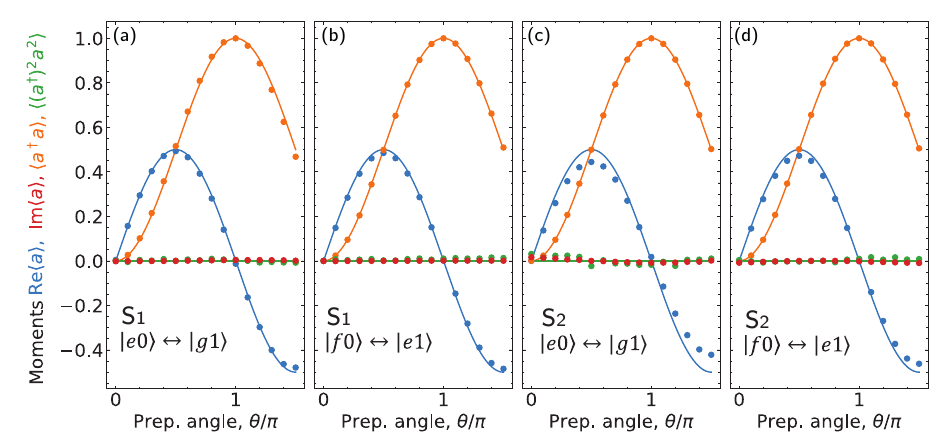}
    \caption{\textbf{Single photon moments}. Moments of an emitted photonic field as a function of the preparation angle $\theta$ of the source qubit prior to emission for all source-emitter parametric gates used. (a) $\ket{e0} \leftrightarrow \ket{g1}$ on $\mathrm{S}_1$,  (b) $\ket{f0} \leftrightarrow \ket{e1}$ on $\mathrm{S}_1$,  (c) $\ket{e0} \leftrightarrow \ket{g1}$ on $\mathrm{S}_2$,  (d) $\ket{f0} \leftrightarrow \ket{e1}$ on $\mathrm{S}_2$.
    }
    \label{fig:appendix_g2}
\end{figure*}

This is a system of coupled linear equations which can be solved to extract the moments $\langle (a^{\dagger})^{s} a^{t} \rangle$ of the photon. We thus take the complex $[I,Q]$ values measured for each experimental shot and extract the moments $\langle (a^{\dagger})^{s} a^{t} \rangle$ by solving this linear system, truncating at order $s,t = 2$. We can test this procedure, the chosen truncation and the process of photon emission by preparing a source qubit in the ground state, rotating by an angle $\theta$, then performing an $\ket{e0} \leftrightarrow \ket{g1}$ (or $\ket{f0} \leftrightarrow \ket{e1}$) emission and measuring the resulting output field. Figure \ref{fig:appendix_g2} shows the reconstructed moments of a photonic field as a function of $\theta$ for such an experiment, performed using $\ket{e0} \leftrightarrow \ket{g1}$ (a) and $\ket{f0} \leftrightarrow \ket{e1}$ (b) pulses on qubit $\mathrm{S}_1$ and $\ket{e0} \leftrightarrow \ket{g1}$ (c) and $\ket{f0} \leftrightarrow \ket{e1}$ (d) pulses on qubit $\mathrm{S}_2$. 
\label{Appendix:MPS_tomography}
We see that the measured $\langle a \rangle$ moments oscillate sinusoidally, reaching 0 at $\theta = \pi$, while the moment $\langle a^\dagger a \rangle$ reaches a maximum at this point, closely following the expected moments for a state of form $\ket{\psi}=\cos(\theta)\ket{g}+\sin(\theta)\ket{e}$. In addition, we observe that  $\langle (a^\dagger)^2 a^2 \rangle$ remnains close to zero throughout. This indicates that the emission is a single photon process, i.e. the second order correlation coefficient $g^{(2)}=\langle \hat{a}^{\dagger 2} \hat{a}^2 \rangle/\langle \hat{a}^{\dagger} \hat{a} \rangle^2 \approx 0$. This also confirms that truncation at $s,t = 2$ is a valid choice. This method, however, relies on being able to translate $[I,Q]$ into the phase space of the photonic field. This amounts to scaling the $I$ and $Q$ axes correctly. To extract a scale, we assume that the field we measure after preparing $S_1$ in the state $\ket{e}$ and performing an $\ket{e0} \leftrightarrow \ket{g1}$ emission pulse corresponds to exactly one photon. By choosing this scale, we have implicitly calibrated out any photon losses between the chip and the detector, which is indistinguishable from added noise.

\begin{figure*}
    \centering
    \includegraphics[width=0.90\textwidth]{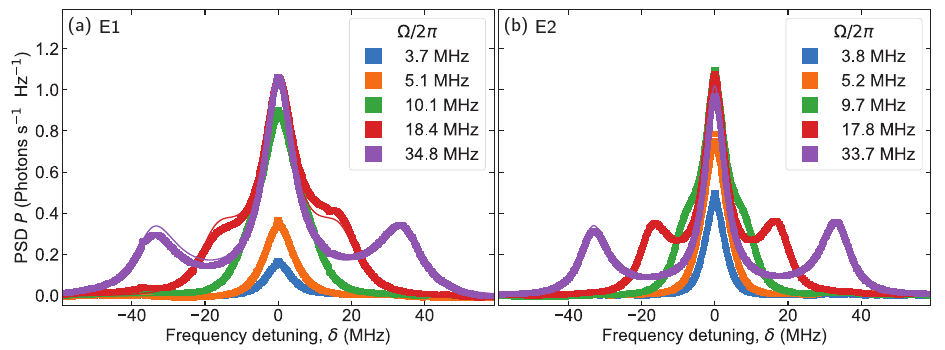}
    \caption{\textbf{Mollow triplet spectroscopy}. Power spectral density (PSD) spectra were measured for $\mathrm{E}_1$ (a) and $\mathrm{E}_2$ (b) at a range of Rabi drive frequencies $\Omega$. The transition from the unsaturated (blue) to fully saturated (purple) photon blockade regime is clearly visible in the PSD spectrum as a splitting of the single qubit peak into the characteristic Mollow triplet lineshape.}
    \label{fig:appendix_Mollow}
\end{figure*}

To further confirm the validity of the extracted scaling factor, we perform a gate-independent measurement that does not rely on high fidelity of single qubit preparation or the $\ket{e0} \leftrightarrow \ket{g1}$ or $\ket{f0} \leftrightarrow \ket{e1}$ emission pulses. Instead, we simply drive the emitter qubit $\mathrm{E}_1$ with a continuous tone via its drive line and measure the transmitted power spectral density to the emission line as a function of frequency for different drive powers, as shown in Fig.~\ref{fig:appendix_Mollow}. Far away from resonance, the emitter suppresses transmission, but close to resonance the qubit is driven and permits power to pass through to the emission line. At low drive powers (blue), we observe a single peak, corresponding to a weakly driven emitter qubit whose linewidth is determined primarily by the strong coupling to the emission line. As we increase the power to a level where the Rabi frequency $\Omega$ exceeds the bandwidth of the qubit, we see the single peak split into the characteristic Mollow triplet~\cite{Mollow1969}. This splitting indicates we are in the so-called photon blockade regime. This allows us to fit the spectrum and extract a value for the scaling factor~\cite{Besse2020a,Reuer2022}. The value we obtain agrees with the scale extracted from a Fock state $\ket{1}$ emitted via the $\ket{e0} \leftrightarrow \ket{g1}$ emission to within 4 \%. The remaining discrepancy may be due to the presence of an additional emitter qubit coupled via the common output line, which may explain the small deviation of the measured Mollow triplet from the fitted model.

\section{Reconstruction of large matrix product operators}

To reconstruct up to 4 mode photonic states, we extract $[I,Q]$ values for each photonic mode and each shot. We then calculate the joint 4 photon moments $\langle \hat{a}_1^{\dagger s} \hat{a}_1^{t} \hat{a}_2^{\dagger u} \hat{a}_2^{v} \hat{a}_3^{\dagger w} \hat{a}_3^{x} \hat{a}_4^{\dagger y} \hat{a}_4^{z} \rangle$, where $s,t... ,z \in \{0,1\}$, of all constituent photonic modes P$_i$. To find the most likely physical density matrix $\rho$, we then perform a maximum likelihood estimation using the means and variances of the extracted moments~\cite{Eichler2014, Ferreira2024} using the convex optimization python package CVXPY~\cite{Diamond2016,Agrawal2018}. We choose the measurement basis for each photon P$_i$ such that the entry $\rho(0,2^{i-1})$ becomes real. To correct for small frequency drifts in the emitter qubit frequencies, we re-calibrate the parametric drive frequencies every $\sim 4$ hours.
Directly reconstructing density matrices for significantly more than 4 modes becomes extremely difficult, as the required number of measurements to accurately extract moments of order $O$ scales with $\eta^{O}$~\cite{daSilva2010}, where $\eta \approx 0.25$. 

However, the cluster state is the unique ground state of a gapped Hamiltonian with only nearest-neighbour terms~\cite{Raussendorf2003}: 
\begin{equation}
\label{eq:cluster_state_hamiltonian}
    \hat{H} = -\sum_i \hat{X}_{i} \prod_{j \in \mathbf{N}(i)} \hat{Z}_j
\end{equation}
where $\mathbf{N}(i)$ indicates all nearest-neighbours of the photon P$_i$, and $\hat{X},\hat{Z}$ are Pauli operators. 

Measuring the Hamiltonian can be used to certify the state~\cite{Cramer2010}, however, this does not take into account all of the information from the reduced density matrices. Thus, we employ the maximum likelihood estimation in~\cite{Baumgratz2013} to reconstruct the maximum likelihood
state given the measured reduced density matrices. As  supports for the reduced density matrices we use the same supports as the terms in the cluster Hamiltonian~\ref{eq:cluster_state_hamiltonian}, which guarantees that in the limit of perfect preparation and measurement, the cluster state is uniquely determined by these reduced density matrices. Given a $2\times n$ photonic qubit cluster state, we only need to reconstruct local density matrices $\rho_i$ of nearest-neighbour qubits, which in our case consist of a maximum of 4 photonic modes. We then perform maximum likelihood estimation on the $\rho_i$ to find the most likely physical full density matrix $\rho$~\cite{Baumgratz2013, Lanyon2017}. In practice, while the local density matrices $\rho_i$ at the ends of the ladder-like structure consist of only 3 photonic modes, we choose to use a single 4 photonic mode density matrix at the edges to simplify the analysis procedure. For consistency, we choose one measurement basis for each photon for all local density matrices $\rho_i$, by ensuring that $\rho_i(0,2^{i-1}) \in \mathbb{R}$.
For the local density matrices $\rho_i$ the highest order moments are of order $O\leq4$, independent of the size of the full cluster state. Therefore, this procedure allows us to reconstruct arbitrarily large photonic states with a total number of shots that does not increase beyond that required to measure moments of order $O=4$. 

In addition, to reduce the computational burden of dealing with the ever-growing density matrices, we use a matrix product operator approach to efficiently reconstruct and describe the full density matrices~\cite{Baumgratz2013}. While the ideal cluster has only a bond dimension of 2, the bond dimension needed to reconstruct the state is larger and we find there is a correlation between the reconstructed fidelities of experimentally generated states and the truncation used to limit the bond dimension of the matrix product operator. We choose a small truncation to maintain an unbiased reconstruction, resulting in bond dimensions $\sim 200$. Even smaller truncation could possibly further increase the accuracy, but also requires higher computational effort. For the reconstruction of the most likely physical matrix product operator, we use the Julia package iTensor~\cite{Fishman2022} and and a graphics processing unit (GPU).

\begin{figure*}
    \centering
    \includegraphics[width=0.95\textwidth]{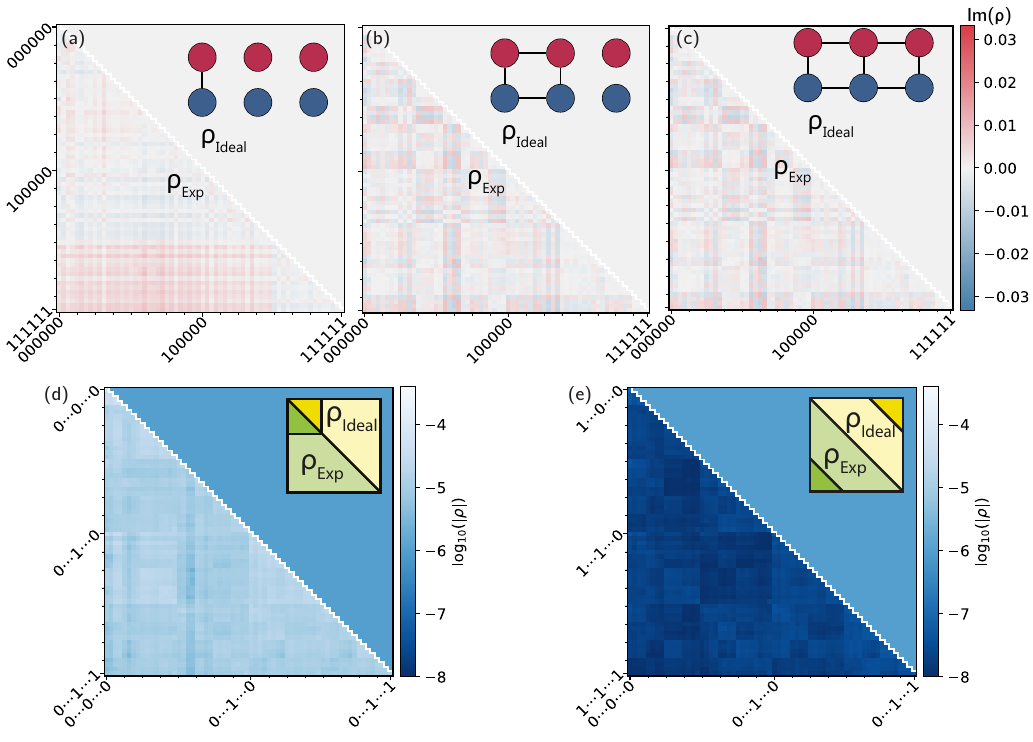}
    \caption{\textbf{Density matrix imaginary components and magnitude.}  For completeness, we show here the imaginary component of the states shown in Fig.~\ref{fig:Cluster_appearance}(b),(d),(f) and the magnitude of the states shown in Fig.~\ref{fig:rho_2d_states}(a),(b). We keep the same convention from Fig.~\ref{fig:rho_2d_states}, showing the density matrix of the corresponding ideal state in the upper-right corner, and show only subsets of the 20 photon cluster state. The ideal cluster states should all have zero imaginary component and uniform magnitude.}
    \label{fig:appendix_imagdm}
\end{figure*}

\begin{figure}
    \centering
    \includegraphics[width=0.90\columnwidth]{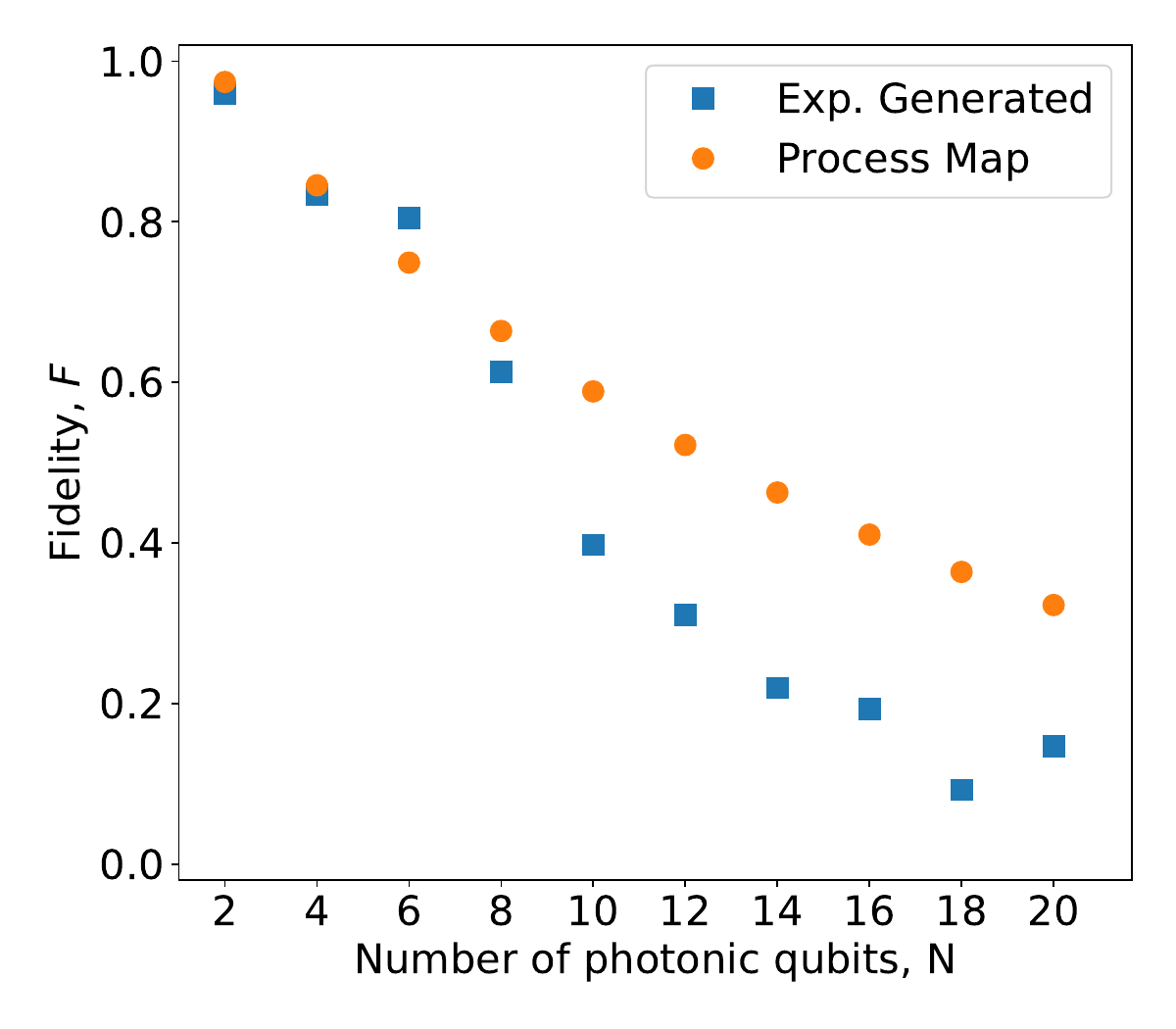}
    \caption{\textbf{Fidelity of reconstructed cluster states}. Fidelity as a function of number of generated photonic qubits $N = 2 \times n$, as measured by matrix product operator tomography of the experimentally generated states (blue, solid squares) and as estimated by process maps (orange, solid circles).}
    \label{fig:fidelity}
\end{figure}

We see the results of this two-step characterization in Fig.~\ref{fig:Cluster_appearance}, where we show the real components of the density matrix as we build up a 6 qubit cluster state (b,d,f), and in Fig.~\ref{fig:rho_2d_states} where we show the phase of two subsets of the density matrix of a 20 qubit state (a,b). We show the imaginary component of the same 6 photonic qubit states in Fig.~\ref{fig:appendix_imagdm}(a,b,c) and the magnitude of the two subsets of the 20 photonic qubit state in Fig.~\ref{fig:appendix_imagdm}(d,e). The fidelity of the reconstructed cluster states is shown in Fig.~\ref{fig:fidelity}.

\begin{figure}
    \centering
    \includegraphics[width=0.90\columnwidth]{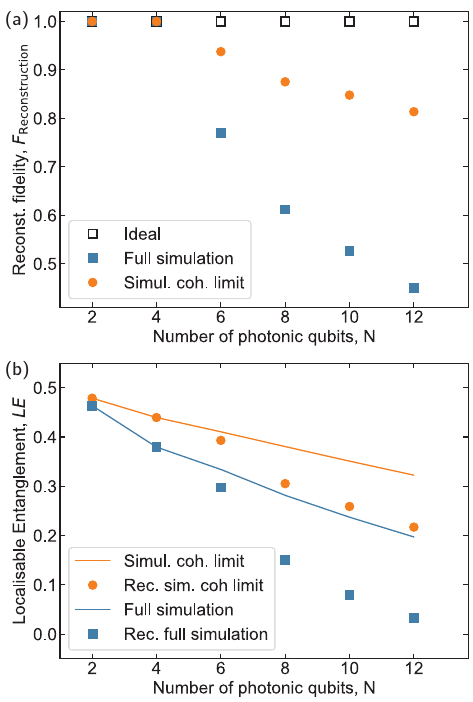}
    \caption{\textbf{Limitation of reconstruction method}. (a) Reconstruction fidelity $F_{\text{Recontruction}}$, defined as the fidelity between the original density matrix and the density matrix obtained after reconstruction, as a function of number of generated photonic qubits $N = 2 \times n$, for the ideal cluster state (black square) and for density matrix obtained by simulations, taking into account all errors (blue squares) or only decoherence (orange circles) (b) Localisable entanglement of the simulated states taking into account all errors (blue) or only decoherence (orange), before (line) and after reconstruction (dots).
    }
    \label{fig:reconstruction}
\end{figure}

The maximum likelihood estimation underestimates the fidelity of the state relative to the prediction by the process maps. This is due to the fact that for global mixed states, the local reduced density matrices cannot uniquely specify the state as they cannot be used to distinguish between longer range entanglement and statistical mixture. The maximum likelihood estimation does not have any assumptions beyond the provided data and thus converges to a generic state compatible with the measurement record, which will typically underestimate the fidelity relative to a specific state. We confirm this hypothesis by running the estimation on simulated data, shown in Fig.\ref{fig:reconstruction}, which shows the same pattern.

Specifically, we applied the reconstruction method to simulated density matrices, including different kinds of errors, and compare the reconstructed density matrix to the original one from simulation, see Fig.~\ref{fig:reconstruction}(a). While for cluster states with $N<6$, the fidelity between the reconstructed and the simulated states is close to 1, this fidelity significantly decreases for $N>8$ for all imperfect cluster states. This drop in fidelity is most pronounced for simulated states including all errors, i.e. decoherence, leakage and coherent errors. The drop is smaller for simulations that only consider decoherence. The fidelity between the reconstructed and the simulated states is below 50~\% for $N=12$, when including all errors, and therefore raises the question of how accurate the localizable entanglement values extracted with this method (shown in Fig.~\ref{fig:negativity_and_schematic}) are. To investigate this, we calculate the localizable entanglement of the simulated states before and after applying the reconstruction method, see Fig.~\ref{fig:reconstruction}(b). We find that the reconstruction method consistently results in lower localizable entanglement values, especially for states $N>8$. We can thus assume that the localizable entanglement values obtained via the reconstruction method underestimate the actual values. Therefore, we consider the obtained localizable entanglement values to be a conservative estimate of the actual localizable entanglement.

\begin{figure}
    \centering
    \includegraphics[width=0.90\columnwidth]{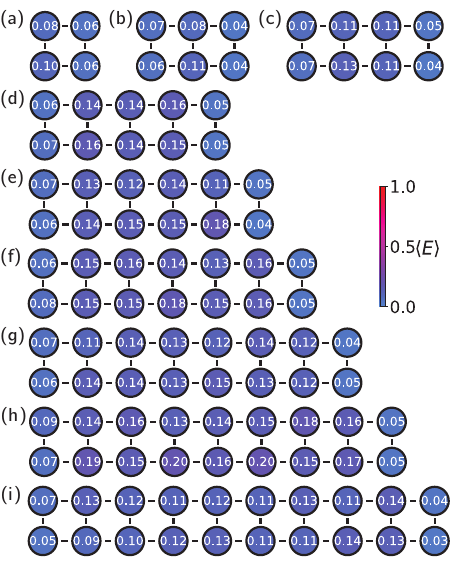}
    \caption{\textbf{Cluster state photon ground state energies}. The energy $\langle E \rangle$, as defined in \ref{eq:cluster_state_energy}, of each photonic qubit for (a) 4, (b) 6, (c) 8, (d) 10, (e) 12, (f) 14, (g) 16, (h) 18 and (i) 20 photonic qubit states. The ideal state is the ground state of the Hamiltonian in \ref{eq:cluster_state_hamiltonian} and has $E_i = 0$.}
    \label{fig:ground_state_energies}
\end{figure}

In addition, we can characterize the proximity of the global state of the constituent qubits to those of the target state by estimating the local energies 
\begin{equation}
\label{eq:cluster_state_energy}
    E_i =-\frac{1}{2} \langle \hat{X}_{i} \prod_{j \in \mathbf{N}(i)} \hat{Z}_j\rangle +\frac{1}{2}
\end{equation}
of each photon P$_i$ from the local four-qubit density matrices. For the ground state, i.e. the ideal cluster state, all local energies $E_i$ are zero, while for the first excited state of the Hamiltonian $E_i=1$. For the generated $4$ to $20$ photonic qubit cluster states, we obtain energies $E_i$ of $0.06-0.08$ for first and last pair emitted photonic qubits, and $0.12-0.26$ for the other photonic qubits, independent of the size of the generated states, as shown in Fig.~\ref{fig:ground_state_energies}(a-h). This demonstrates our ability to generate an arbitrary size $2 \times n$ cluster state with a constant error rate per emission step. The first two and last two photonic qubits to be emitted have only two entanglement bonds each (as opposed to three) and thus have fewer gates applied to them. The fact that these qubits consistently have the lowest ground state energies suggests that the finite fidelity of the entangling gates is a significant source of error in our protocol.

\section{Cluster state generation protocol}
\begin{figure}
    \centering
    \includegraphics[width=0.48\textwidth]{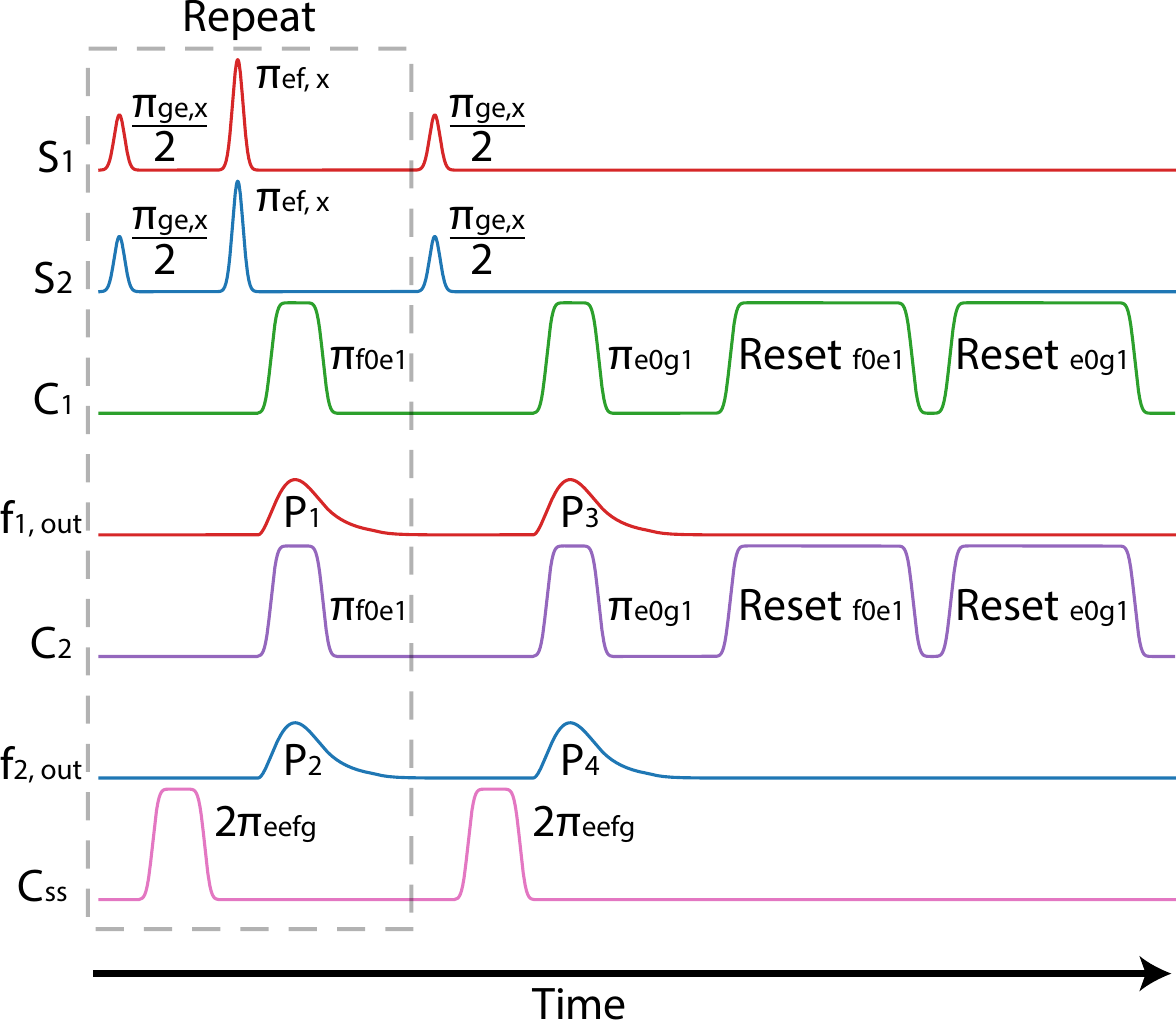}
    \caption{\textbf{Cluster state pulse sequence.} Pulses sent to each element of the device during the generation of an N-photon cluster state. The pulses in the dashed box are repeated $N/2-1$ times. $f_{1, out}$ and $f_{2, out}$ denote the field in the output line at frequencies $f_1, f_2$ corresponding to the frequencies of the emitter qubits, $\mathrm{E}_1$ and $\mathrm{E}_2$, respectively. Pulses are not to scale.}
    \label{fig:pulse_sequence_cluster4}
\end{figure}

An N-qubit cluster state is generated by first preparing all qubits in the $\frac{1}{\sqrt{2}}\left(\ket{g}+\ket{e}\right)$ superposition state, then applying CPHASE gates between all neighbouring qubits. The gate sequence depicted in Fig.~\ref{fig:layout}(f) realises this for the ladder-like entanglement structure depicted in Fig.~\ref{fig:layout}(a) using the device's native gate set. We perform single qubit gates using \SI{128}{\nano\second} duration gaussian envelope DRAG pulses~\cite{Motzoi2009}. Two-qubit gates are perfomed by generating smooth, flat-topped pulses utilizing the error function for the shape of the rising and falling edges. As the frequency of these pulses is in the sub-GHz range, the waveforms may be directly digitally synthesized without the need for up-conversion. Dispersive qubit readout is performed using frequency up-converted square pulses with a duration of \SI{1.2}{\micro\second}. Source qubit reset is achieved via a \SI{7}{\micro\second}  $\ket{f0} \leftrightarrow \ket{e1}$ flat-topped pulse followed by an \SI{8.5}{\micro\second} $\ket{e0} \leftrightarrow \ket{g1}$ flat-topped pulse on both source qubits. This ensures any residual source qubit population in the $\ket{e}$ or $\ket{f}$ states is transferred into the emitters and discarded, resetting the source qubits into $\ket{g}$. This fast qubit reset allows us to perform an entire cluster state generation protocol for up to 20 photonic qubits with a \SI{32}{\micro\second} repetition time. The full sequence for generation of a cluster state is shown in Fig.\ref{fig:pulse_sequence_cluster4}. To generate an $N$ photonic qubit state, the pulses within the dashed box are repeated $N/2 - 1$ times. 

\section{Process tomography}
\label{appendix:process_maps}
To further characterize the generated cluster state, we perform process tomography on two different emission processes: The first process $p_1$ includes a Hadamard gate on both source qubits $\mathrm{S}_1$ and $\mathrm{S}_2$, followed by a CPHASE gate between $\mathrm{S}_1$ and $\mathrm{S}_2$ and a controlled emission via a CNOT gate from $\mathrm{S}_1$ and $\mathrm{S}_2$. This is the process used repeatedly to emit the first $N-2$ photonic qubits. For the final emission, the CNOT gates on $\mathrm{S}_1$ and $\mathrm{S}_2$ are replaced with a SWAP to disentangle the photonic qubits from the source qubits. The second process $p_2$ we characterize thus consists of a Hadamard on $\mathrm{S}_1$ and $\mathrm{S}_2$, followed by a CPHASE between $\mathrm{S}_1$ and $\mathrm{S}_2$ and unconditional emission via a SWAP gate from $\mathrm{S}_1$ and $\mathrm{S}_2$.

The process map describing these processes $\chi: \rho_{S_1,S_2}^{\text{pre}} \rightarrow \rho_{S_1,S_2,P_1,P_2}^{\text{post}}$ maps the density matrix of the two source qubits $\mathrm{S}_1$ and $\mathrm{S}_2$ before the process $\rho_{S_1,S_2}^{\text{pre}}$ to the joint density matrix $\rho_{S_1,S_2,P_1,P_2}^{\text{post}}$ of the source qubits $\mathrm{S}_1$ and $\mathrm{S}_2$ and the generated photonic qubits P$_1$ and P$_2$ after the process. In the Pauli basis, we can write

\begin{align}
\rho_{S_1,S_2}^{\text{pre}} &= \sum_{m,n} \rho_{m,n}^{\text{pre}} \hat{\sigma}_m \otimes \hat{\sigma}_n \\
\rho_{S_1,S_2,P_1,P_2}^{\text{post}} &= \sum_{i,j,k,l} \rho_{i,j,k,l}^{\text{post}} \hat{\sigma}_i \otimes \hat{\sigma}_j \otimes \hat{\sigma}_k \otimes \hat{\sigma}_l\\
&= \chi \rho_{S_1,S_2}^{\text{pre}} \\
&= \sum_{i,j,k,l} \left(\sum_{m,n} \chi_{i,j,k,l}^{m,n} \rho_{m,n}^{\text{pre}} \right) \hat{\sigma}_i \otimes \hat{\sigma}_j \otimes \hat{\sigma}_k \otimes \hat{\sigma}_l \label{eq:process_def}
\end{align}

To measure the process map $\chi$, we prepare the source qubits in 16 different initial states
\begin{align}
\{\ket{g},\frac{1}{\sqrt{2}} \left( \ket{g} + \ket{e} \right),\frac{1}{\sqrt{2}} \left( \ket{g} - i \ket{e} \right), \ket{e} \}^{\otimes2},
\end{align}
apply the process to each intial state, and perform a state tomography to measure $\rho_{S_1,S_2,P_1,P_2}^{\text{post}}$ for each initial state. We thus need to perform a joint state tomography on stationary and photonic qubits. To do this, we measure the source qubits $\mathrm{S}_1$ and $\mathrm{S}_2$ after the process is completed in $\mathbf{B}=X, Y, Z$ basis by optionally applying a $\pi/2$ rotation around the $X$ or $Y$ axis and reading out both source qubits. In addition, we detect the generated photonic qubits as described in App.~\ref{Appendix:experimental_setup}.

The readout and photon traces are integrated with filters (see App.~\ref{Appendix:Photon_transients}), such that for each shot, we obtain 4 pairs of $[I,Q]$ values, one from each qubit. We then assign the most likely state $s_1$/$s_2$ ($\ket{g}$,$\ket{e}$,$\ket{f}$) to each of the readout traces based on the $[I,Q]$ value and the estimated tri-modal Gaussian distribution~\cite{Magnard2018,Reuer2023}. We observe that the measured population outside the computational space ($\ket{g}$,$\ket{e}$) is below 1~\%. We can therefore safely discard measurement outcomes where a qubit is measured in the $\ket{f}$ state without significantly affecting the reconstructed process map and subsequently build a joint histogram $H(I_1,Q_1,I_2,Q_2,s_1,s_2)_{\mathbf{B_1},\mathbf{B_2}}$ where $I_1$ and $Q_1$ refers to the measured $[I,Q]$ values from P$_1$ and $I_2$ and $Q_2$ refer to P$_2$ for each of the 9 joint measurement bases $(\mathbf{B_1},\mathbf{B_2})$. 

We reconstruct the two-photon density matrices $\rho(s_1,s_2)_{\mathbf{B_1},\mathbf{B_2}}$ for each of the 4 possible qubit states and 9 possible measurement bases from the histogram using the methods described in Ref.~\cite{Eichler2012,Besse2020a,Reuer2022}. We then calculate the joint four-qubit POVM expectation values:
\begin{align}
\langle \Pi_{B_1,B_2,B_3,B_4}^{ijkl} \rangle &=  H(i,j)_{\mathbf{B_1},\mathbf{B_2}} \bra{\Pi_{B_3,B_4}^{kl}} \rho(i,j)_{\mathbf{B_1},\mathbf{B_2}} \ket{\Pi_{B_3,B_4}^{kl}}
\end{align}
where
\begin{align}
H(i,j)_{\mathbf{B_1},\mathbf{B_2}}&=\sum_{I_1,Q_1,I_2,Q_2} H(I_1,Q_1,I_2,Q_2,s_1,s_2)_{\mathbf{B_1},\mathbf{B_2}}
\end{align}
is the reduced histogram, $\Pi_{B_1,B_2,B_3,B_4}^{ijkl}$ is the four qubit POVM operator for measurements in the bases $(B_1,B_2,B_3,B_4)$ with outcomes $ijkl$, and $\Pi_{B_3,B_4}^{kl}$ is the POVM operator on the photonic qubits.

\begin{figure}
    \centering
    \includegraphics[width=0.48\textwidth]{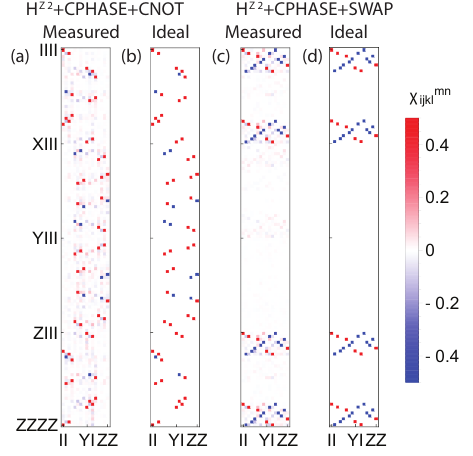}
    \caption{\textbf{Process Maps.} Measured (a,c) and ideal (b,d) process maps $\chi_{ijkl}^{mn}$ in Pauli basis (see Eq.~\ref{eq:process_def}) of (a,b) the process $p_1$ with H$^{\otimes 2}$-CPHASE-CNOT$^{\otimes 2}$ used for the first $N-1$ emission steps, and (c,d) and the final emission process $p_2$ with H$^{\otimes 2}$-CPHASE-SWAP$^{\otimes 2}$.}
    \label{fig:ProcessMaps}
\end{figure}

From the four-qubit POVM expectation value, we reconstruct the most likely physical density matrix using the maximum-likelihood method introduced in Ref.~\cite{Hradil2004}. Obtaining a four-qubit density matrix for all 16 initial states, we, using Eq.~\ref{eq:process_def}, obtain a system of linear equations. Solving this system, we obtain the process maps $\chi_{p_1}$ and $\chi_{p_2}$ shown in Fig.~\ref{fig:ProcessMaps} for the two processes.

To verify that the inferred process map is physical, i.e. trace preserving, we calculate the Choi matrix $C^{\chi}$~\cite{Choi1975, Schwartz2016, Besse2020a} from the process map $\chi$:
\begin{align}
C^{\chi} &= \sum_{ijkl} \hat{e}_{ijkl} \otimes  \chi(\hat{e}_{ijkl})
\end{align}
where $\hat{e}_{ijkl} \in \mathcal{C}^{2\times 2\times 2 \times 2}$ is a unit tensor with 1 as the $ijkl$-th entry and 0 otherwise. Note that $\chi_{ijkl}^{mn}$ (see Fig.~\ref{fig:ProcessMaps}) gives the process map in the Pauli basis, while for the Choi matrix we consider the map in the $\hat{e}_{ijkl}$ basis. The process map $\chi$ is trace-preserving if and only if the Choi matrix $C^{\chi}$ is positive, which is the case if
\begin{align}
\text{Tr}\left[C^{\chi} \hat{\sigma}_i \otimes \hat{\sigma}_j  \otimes \hat{\sigma}_0 \otimes \hat{\sigma}_0 \otimes \hat{\sigma}_0 \otimes \hat{\sigma}_0\right] &= 4 \delta_{i0}\delta_{j0}, \label{eq:tracepreseve}
\end{align}
for $i,j=0,1,2,3$, where $\hat{\sigma}_0$ is the identity operator and $\hat{\sigma}_1,\hat{\sigma}_2,\hat{\sigma}_3$ are the other Pauli operators. We find that Eq.~\ref{eq:tracepreseve} is fulfilled for all inferred process maps, indicating that the reconstructed process maps are physical.

As the Choi matrix is a positve matrix, we can calculate the fidelity to the ideal Choi matrix $F=\sqrt{C^{\chi}_{\text{ideal}}}\sqrt{C^{\chi}_{\text{exp}}}$ and find fidelities of $F=87 \%$ for the process $p_1$ involving controlled emission (Fig.~\ref{fig:ProcessMaps}(a,b)) and $F=89 \%$ for the final emission step $p_2$(Fig.~\ref{fig:ProcessMaps}(c,d)).

With the measured process maps, we directly obtain a matrix product operator representation of the generated cluster state, assuming the first $2 \times (n-1)$ sites to be described by the operator corresponding to $p_1$ process map $\chi_{p_1}$ and the last 2 sites to be described by the operator corresponding $p_2$ process map $\chi_{p_2}$. Contracting the appropriate indices, we can extract the predicted fidelities and the predicted amount of localisable entanglement.

During the process tomography, we need to measure two-photonic-qubit density matrices for 16 different initial states, 9 two-qubit measurement bases and 4 possible qubit states. Consequently, a large amount of measurement shots is required and we needed about 48 hours to acquire the data for each process map. During these 48 hours, the phase of the room temperature measurement electronics experiences a drift. During the process tomography, we repeatedly measured these drifts using two $1/\sqrt{2}(\ket{g}+\ket{e})$ photonic qubits as a reference. We then correct for these phase drifts by rotating the inferred two-photonic-qubit density matrices from the corresponding time period.

\section{Master equation simulation}
\label{appendix:simulation_details}
Our full master equation simulations are based on the quantum circuit description of the cluster state emission protocol, see Fig~\ref{fig:layout}f. We consider two source qutrits and a register of $2\times n$ qubits where the cluster state will be prepared. The emitter transmons are not considered, thus controlled emissions are executed between one of the source qutrits and the photonic qubit of $j$-th emission cycle. For a given cluster state size we simulate $n-1$ Hadamard-CPHASE-CNOT emission cycles and one Hadamard-CPHASE-SWAP emission cycle. Since the qubit register is a representation of the photonic state, it is not subject to the action of any noise processes. On the other hand, we account for  decoherence, coherent errors, and leakage of the source qutrits. Our simulations were implemented using the open source library Cirq and can be found in~\cite{simulation_repo}.

Decoherence due to finite life and coherence times of both the $|e\rangle$ and $|f\rangle$ of the source qutrits is modelled as amplitude and phase damping, respectively. For the former the qutrit state evolves as $\rho\to\mathcal{E}_{\rm AD}[\rho]$ with the qutrit amplitude damping channel given by
\begin{equation}
\label{eqn:qutrit_ad_channel}
\mathcal{E}_{\rm AD}[\rho] = \sum_{l=1}^3 \hat{M}^{(l)}_{\rm AD} \rho \hat{M}^{(l)}_{\rm AD},
\end{equation}
with Kraus operators
\begin{subequations}
\begin{align}
\hat{M}^{(1)}_{\rm AD} &= 
\begin{pmatrix}
1 & 0 & 0 \\
0 & \sqrt{1 - p^{(eg)}_{\rm AD}}  & 0 \\
0 & 0 & \sqrt{1 - p^{(fe)}_{\rm AD}} 
\end{pmatrix},
\\
\hat{M}^{(2)}_{\rm AD} &= 
\begin{pmatrix}
0 & \sqrt{p^{(eg)}_{\rm AD}} & 0 \\
0 & 0 & 0 \\
0 & 0 & 0 \\
\end{pmatrix}, \enspace
\hat{M}^{(3)}_{\rm AD} = 
\begin{pmatrix}
0 & 0 & 0 \\
0 & 0 & \sqrt{p^{(fe)}_{\rm AD}} \\
0 & 0 & 0 \\
\end{pmatrix},
\end{align}
\end{subequations}
and probabilities $p^{(ge, fe)}_{\rm AD} = 1 - e^{-t/T_1^{(e,f)}}$ where $t$ is the time interval during which the channel acts and $T_1^{(e,f)}$ are the lifetimes of $|e\rangle$ ($|f\rangle$), see Table~\ref{table:parameters}.

Similarly under the action of phase damping the state evolves as $\rho\to\mathcal{E}_{\rm PD}[\rho]$, with the qutrit phase damping channel given by 
\begin{equation}
\label{eqn:qutrit_pd_channel}
\mathcal{E}_{\rm PD}[\rho] = \sum_{l=1}^3 \hat{M}^{(l)}_{\rm PD} \rho \hat{M}^{(l)}_{\rm PD},
\end{equation}
and corresponding Kraus operatos
\begin{subequations}
\begin{align}
\hat{M}^{(1)}_{\rm PD} &= 
\begin{pmatrix}
1 & 0 & 0 \\
0 & \sqrt{1 - p^{(e)}_{\rm PD}}  & 0 \\
0 & 0 & \sqrt{1 - p^{(f)}_{\rm PD}} 
\end{pmatrix},
\\
\hat{M}^{(2)}_{\rm PD} &= 
\begin{pmatrix}
0 & 0 & 0 \\
0 & \sqrt{p^{(e)}_{\rm PD}} & 0 \\
0 & 0 & 0 \\
\end{pmatrix}, \enspace
\hat{M}^{(3)}_{\rm PD} = 
\begin{pmatrix}
0 & 0 & 0 \\
0 & 0 & 0 \\
0 & 0 & \sqrt{p^{(f)}_{\rm PD}} \\
\end{pmatrix},
\end{align}
\end{subequations}
and probabilities $p^{(e, f)}_{\rm PD} = 1 - e^{t/T_1^{(e,f)}}e^{-2t/T_2^{*(e,f)}}$ where $t$ is the time interval during which the channel acts, $T_1^{(e,f)}$ and $T_2^{*(e,f)}$ are the life and coherence times of $|e\rangle$ ($|f\rangle$). We include decoherence in our simulations applying the composed channel $\mathcal{E}_{\rm PD} \circ \mathcal{E}_{\rm AD}$ symmetrically after the action of each single-qutrit, two-qutrit, and qutrit-qubit gates, where the error probabilities are fixed by the respective gate times (see Sec.~\ref{sec:device_and_protocol}). Since the total emission cycle takes $650$ns, the two source qutrits idle for the last $\sim75$ns of the sequence, during which they experience decoherence.

The total error on the single qubit gates is modelled as a combination of decoherence and coherent errors, the latter given by an under-rotation. Specifically, the imperfect Hadamard in the qubit subspace of the source qutrit is a $(\pi-\gamma_{\rm H})$ rotation around the $\frac{X+Z}{\sqrt{2}}$ axis, and the imperfect population transfer between $|e\rangle$ and $|f\rangle$ is a $(\pi-\gamma_{\pi})$ rotation around the $x$-axis. In both cases we ignore a global phase of $i$. The under-rotation angles are obtained based on the results of randomized benchmarking for all Clifford single qubit gates. We find $\gamma_{\rm H}\approx\gamma_{\pi}\approx 0.25^{\circ}$.

\begin{figure*}
    \centering
    \includegraphics[width=0.90\textwidth]{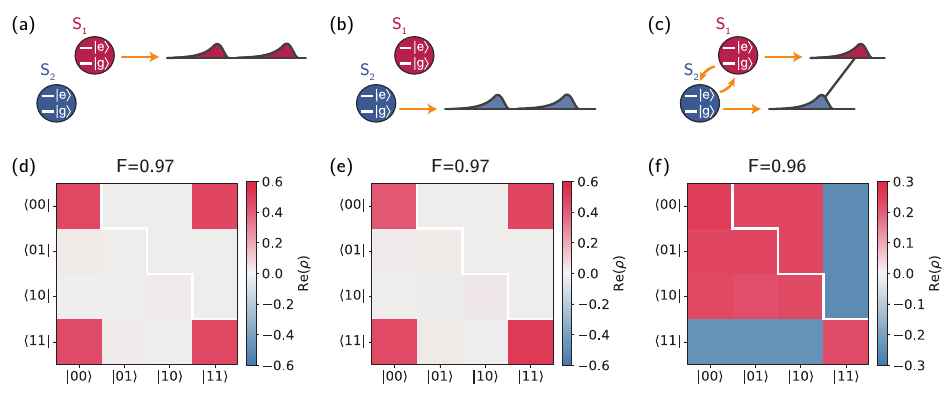}
    \caption{\textbf{Two-qubit entangled states.} (a) Bell state generated via a CNOT gate $(\ket{00}+\ket{11})/\sqrt{2}$, emitted from $\mathrm{S}_1$. (b) Bell state  generated via a CNOT gate $(\ket{00}+\ket{11})/\sqrt{2}$, emitted from $\mathrm{S}_2$. (c) Maximally entangled state generated via a CPHASE gate $(\ket{00}+\ket{10}+\ket{01}-\ket{11})/2$, emitted simultaneously from $\mathrm{S}_1$ and $\mathrm{S}_2$.}
    \label{fig:Bell_states}
\end{figure*}

The total error on the qutrit-qutrit or the qutrit-qubit gates is modelled as a combination of decoherence, coherent error, and leakage. In particular, our model of an imperfect CPHASE is implemented as a $(2\pi-\gamma_{\rm CZ})$ rotation on the $|ee\rangle \leftrightarrow |fg\rangle$ transition, leading to phase accumulation by those two states which does not add to $-1$. Based on the preparation fidelity of the two-qubit cluster state (see Fig.~\ref{fig:Bell_states}(c,f)) we find $\gamma_{\rm CZ}\sim0.5^{\circ}$. Additionally, leakage for this gate is modeled as the exchange~\cite{Varbanov2020,Camps2024}
\begin{subequations}
\begin{align}
|ee\rangle &\rightarrow \sqrt{1 - 4L_{\rm CZ}}|ee\rangle + e^{i\phi}\sqrt{4L_{\rm CZ}} |fg\rangle, \nonumber \\
|fg\rangle &\rightarrow \sqrt{1 - 4L_{\rm CZ}}|fg\rangle - e^{-i\phi}\sqrt{4L_{\rm CZ}} |ee\rangle, \nonumber
\end{align}
\end{subequations}
with $0\le L_{\rm CZ}\le 0.25$ the leakage probability. Our model of an imperfect parametric excitation exchange between the source qutrit and the emitter qubit (photonic qubit in our case), only considers decoherence and leakage. The latter is modelled as the exchange
\begin{subequations}
\begin{align}
|f0\rangle &\rightarrow \sqrt{1 - 4L_{\pi}}|e1\rangle + e^{i\phi}\sqrt{4L_{\pi}} |f0\rangle, \nonumber \\
|e1\rangle &\rightarrow \sqrt{1 - 4L_{\pi}}|f0\rangle - e^{-i\phi}\sqrt{4L_{\pi}} |e1\rangle, \nonumber
\end{align}
\end{subequations}
   with $0\le L_{\pi}\le 0.25$ the leakage probability for this gate. The calibration data suggested values of $L_{\rm CZ} = 0.02$ and $L_{\pi} = 0.01$, which we used in our simulations. We note that our simulations suggest the sequential emission protocol to be insensitive to the leakage phases $\phi$, and in all simulation results shown in the main text we have set them to $\phi=0$. Finally, in all our simulations we have assumed the SWAP gates to be decoherence limited.

\section{Localizable Entanglement}
\label{appendix:LE}


To measure the entanglement for a bi-partite mixed state consisting of two quantum systems $\mathrm{A}$ and $\mathrm{B}$, we can compute a quantity known as the negativity~\cite{Vidal2002}:

\begin{equation}
    N(\rho) = \frac{{|| \rho^{T_A} ||_1 - 1}}{2}
\end{equation}

where $\rho^{T_A}$ is the partial transpose of the state's density matrix $\rho$ with respect to system $\mathrm{A}$ and $|| \rho^{T_A} ||_1$ is the trace norm of $\rho^{T_A}$. This quantity is a measure of the degree of entanglement between the systems $\mathrm{A}$ and $\mathrm{B}$ and is zero for a separable state. To use this as a metric for entanglement of larger entangled states we must utilize a related quantity known as localizable entanglement~\cite{Briegel2001,Hein2006}. This is the maximum entanglement that can be created, or localized, between two constituent qubits of the larger entangled state by performing local measurements on all other qubits. When performing local measurements, we choose either $Z$ or $X$ axis measurements. For graph states, measurement along the $X$ axis may be thought of as bypassing the measured qubit and creating all-to-all entanglement bonds between its neighbours, while a $Z$ axis measurement removes the qubit and all its entanglement bonds from the state. A cluster state consisting of $N$ qubits can be projected into a Bell state consisting of qubits $A$ and $B$ with certainty by performing $X$-axis measurements on a chain of qubits connecting $A$ and $B$ and performing $Z$ axis measurements on all other qubits.  For the ladder-like cluster states generated here, we choose $A$ and $B$ to be the diagonally opposite qubits on the ladder, being separated by the greatest number of entanglement bonds. There are many paths that can be chosen to connect $A$ and $B$, of which one is shown in Fig.~\ref{fig:negativity_and_schematic}. We calculate the projection along all paths of length $n$ where $n = N/2$ is the number of photonic qubits along the time multiplexing axis (as opposed to the frequency multiplexing axis, where we always emit at two different frequencies). For larger states, this corresponds to an impractically large number of possible measurement outcomes to compute. Therefore instead of considering every possible measurement outcome for a given path, we take 1024 randomly sampled measurement outcomes. This results in 1024 projections per path which, for an ideal cluster state, should result in 1024 Bell states. We then calculate the negativity for each of these outcomes and for all length-$n$ paths and take the average value. This quantity is the localizable entanglement shown in Fig.~\ref{fig:negativity_and_schematic}. The ideal cluster state has the maximum possible localizable entanglement value of 0.5; we see that smaller states approach this value but this falls off as more gates are added to create larger states. Provided the value is non-zero to within error we can conclude that some entanglement persists between diagonally opposite qubits. The largest such state for which we measure localizable entanglement greater than zero for all possible paths is $N=20$. As the calculation of localization entanglement becomes computationally heavy for states with $N>10$, we use the Julia package iTensor \cite{Fishman2022} on the cluster computing system Euler of ETH Zurich.

\section{Characterizing other entangled states}

\label{Appendix:Other_entangled_states}
By entangling and emitting two photons we can get a better picture of the fidelity of the different entangling gates and the associated emission process. In Fig.~\ref{fig:Bell_states} we created and measured three maximally entangled two-qubit states consisting of two photons emitted from $\mathrm{S}_1$ (a), $\mathrm{S}_2$ (b) using CNOT gates and a state entangled using a CPHASE gate between $\mathrm{S}_1-\mathrm{S}_2$ followed by simultaneous emission via $\mathrm{E}_1$ and $\mathrm{E}_2$ (c). The result of (a) and (b) is a state $\left(\ket{00}+\ket{11}\right)/\sqrt{2}$, while for (c) we generate a state of the form $\left(\ket{00}+\ket{01}+\ket{10}-\ket{11}\right)/2$, which, to within single qubit rotations, is equivalent to the states generated by the CNOT gate. We obtain fidelities of 97~\% for the Bell states emitted from $\mathrm{S}_1$ and $\mathrm{S}_2$, respectively, and 96 \% for the $\mathrm{S}_1-\mathrm{S}_2$ entangled state.


\end{appendix}

\bibliography{bibliography}

\end{document}